# Electromechanical actuation of pristine graphene and graphene oxide: origin, optimization, and comparison

Jefferson Z. Liu[*], and Jeffrey Hughes

Email: zhe.liu@unimelb.edu.au

Department of Mechanical Engineering, The University of Melbourne, Parkville, VIC 3010, Australia

**Abstract**

It is well recognized that the miniaturization of electromechanical devices will bring a revolution to humanity in the coming decades synonymous with the effects of miniaturizing electronic devices in those previous. An electromechanical actuator — a device that converts electrical energy to mechanical deformation or motion — is the core component of many such devices. Consequently, research interrogating mili-, micro-, and nano-actuation has, and will continue to become increasingly essential. The challenge is that behaviour of actuators at small size scales vastly differs to those at the macroscale. We cannot simply shrink the size of conventional actuators at the macroscale down to the micro/nanoscale. In addition, conventional actuation materials (such as piezoelectric ceramics and shape memory alloys) have poor properties and performances when fabricated at small length scale. There is an urgent need to discover novel actuation materials at small length scale. This paper will review recent advances of graphene-based actuation materials. We will focus on different actuation physical mechanisms of this most well-known two-dimension material. The in-depth physical understanding and insights will lay the ground for further optimization/development of graphene-based actuators. They may also provide valuable knowledge for the design and development of other two-dimensional actuation materials and actuators.

**Introduction**

In the final quarter of the twentieth century, a unique paradigm shift was experienced amongst the medical community. Discovered by therapeutic practitioners, the adverse effects to patients undergoing surgery were frequently caused by intraoperative trauma when gaining access to the treatment site, rather than the treatment itself. This new understanding motivated a transition from traditional large-incision techniques to practices that minimised postoperative trauma, giving birth to the field of minimally invasive surgery (MIS), so coined by Wickham in 1987 [1]. Within a single decade, the medical community had embraced MIS, with over 50 percent of therapeutic procedures drawing on this philosophy [2].

MIS has yielded a myriad of advantages since its inception for patients, surgeons, therapeutic insurance agencies and taxpayers alike. Studies have shown that MIS results in reduced convalescence, decreased pain, smaller incisions and thus better cosmesis, less risk of infection, better postoperative immune function and lower rates of morbidity and mortality, compared with traditional cut-and-sew techniques [1, 3–6]. However, MIS is not without flaws, exposing some prominent limitations of which many are mechanical in nature. These include: the inability to catheterise aneurysms accounting for 38% of failed operations, and so on [3, 7]. Practitioners have unanimously identified that significant improvements need to be made in dexterity, control and effectiveness of MIS operative tooling [2], which has been the principal goal of micro-robotic MIS research over the past decade.

A significant shortfall of current micro-robotic MIS equipment is that the majority of the operative tooling must remain outside the patient, thereby severely compromising the surgeon's ability to reach locations deep within the body. This is predominantly due to the limitations in size and versatility of current micro-actuator design, reducing its feasibility to enter the patient via small incisions made in order to conduct MIS. The success of micro-robotic MIS development therefore hinges on the advancement of micro-actuation technology.

More broadly, it is well recognized that the miniaturization of electromechanical devices will bring a revolution to humanity in the coming decades synonymous with the effects of miniaturizing electronic devices in those previous. They promise and deliver a myriad of applications within industry, including those within the automotive, electronic, aerospace, environmental, and defence [8]. Devices ranging from micro/nanoscale resonators, switches, and valves have applications in tasks as diverse as information processing, molecular manipulation, and sensing. An electromechanical actuator — a device that converts electrical energy to mechanical deformation or motion — is the core component of many such devices. Consequently, research interrogating mili-, micro-, and nano-actuation has, and will continue to become increasingly essential.

The drive for smaller actuators has led to various proposed methods of generating forces and displacements at the millimeter, micrometer, and nanometer scales. Many attempts have been made to simply shrink traditional actuators and motors, from which it has been concluded the behaviour of actuators at small size scales vastly differs to those at the macroscale. Table 1 briefly summarizes various actuation schemes available to micro-device designers, and their inherent traits. In the following, we will provide a short overview of these different schemes and their potential applications at the small size scale.

The electromagnetic actuator, found today in countless products ranging from industrial machinery to consumer electronics, achieved widespread use due to its exceptional energy density and longevity [8]. However, attempts to miniaturize this actuator for milliscale applications have resulted in low output forces/torques [9, 10]. This is due to the poor scalability of the electromagnetic force, scaling as a function of the characteristic actuator length to the fourth power ($F \sim L^4$) [11]. As such, electromagnetic actuators and motors are unsuitable for applications requiring nano, micro and low milli-scale actuation, such as micro-robotic MIS, and resonators. Subsequently other methods of generating an electromechanical response at these very small scales are in keen demand.

Electrostatic force – the attraction/repulsion between two charged objects – offers an actuation mechanism that could potentially address this need. Unlike the electromagnetic actuator, electrostatic actuation improves with a reduction in size, scaling with the inverse-square of the characteristic length ($F \sim L^{-2}$) [8]. This excellent scalability coupled with their intrinsic simplicity has allowed experimental demonstration at and below sizes of 100 μm [12–14]. However, despite their excellent scalability, electrostatic actuators still have fundamentally low output forces/torques. In addition, electrostatic forces are nonlinear in nature, necessitating extremely complex control systems to generate accurate position incrementing. They also suffer from a phenomenon know as snap-down, whereby oppositely charged surfaces short-out and weld together, rendering the actuator inoperable.

The thermal actuator, operating due to the thermo-mechanical response of a material (*i.e.*, mechanical deformation caused by a change in temperature), has proven to be very useful for select applications within the micro-device industry [15, 16]. The scaling law for thermal actuators predict high force/torque output that is directly proportional to the characteristic length ($F \sim L$). Thermal actuation prescribes a more efficient output when compared to electromagnetic [8]. Common shortcomings of these actuators include slow response times and short service lives, as they degrade over time under the applied strain. Hence, these actuators are most suited to microscopic applications requiring infrequent use and high output loads.

Piezoelectric actuation is a phenomenon observed in certain materials that generate a mechanical stress/strain in response to an applied electric field [17]. Research into piezoelectric actuators has revealed many desirable properties, including short response times, linear field-displacement behaviour, negligible backlash and direct drive capability [8]. In addition, the force/torque of a piezoelectric actuator scales directly with the characteristic length (F ~ L) [8]. The main trade-offs of piezoelectric are the small actuation strain (0.01-0.1%) and high operation electric field (kV).

In addition to these alternative actuation principles, there exist other options. These include magnetostrictive (elongates under the application of a magnetic field) [18, 19], osmotic (diaphragm displaces in response to a Venturi-like osmotic fluid pressure) [20, 21], electro-rheological (dielectrically polarised particles within a rheological fluid or gel medium displace upon the application of an electric field) [22, 23], electro-conjugate (dielectric fluid jetting in response to the application of a high DC voltage, which drives a submersed turbine) [24, 25], and opto-mechanical (displaces in response to incident photonic energy) [26, 27]. Whilst each of these principles carry merit and have demonstrated potential for operating under certain conditions in certain applications, they are commonly limited by several factors including low output force/torque, difficulty of fabrication at small scales, limited scope for controllability and slow response times, to name a few.

**Table 1** Summary of the various actuation schemes available to micro-device designers and their inherent properties. See text for details.

| Actuation Scheme | Force/Torque Scaling | Response Speed | Simplicity | Controllability |
|---|---|---|---|---|
| **Electromagnetic** | ~$L^4$ | Fast | Complex | Excellent |
| **Electrostatic** | ~$L^{-2}$ | Fast | Moderate | Moderate |
| **Thermal** | ~$L$ | Slow | Moderate | Poor |
| **Magnetostrictive** | ~$L^4$ | Moderate | Complex | Moderate |
| **Osmotic** | Complex | Slow | Complex | Poor |
| **Piezoelectric** | ~L | Fast | Simple | Excellent |
| **Electro-active materials** | ~L | Moderate-Fast | Simple | Excellent |

The cutting edge in actuation discovery has revealed electro-active (EA) materials as a promising and underdeveloped actuation principle worthy of further investigation. EA materials generate mechanical deformation or motion when charged (doped) with electrons or holes [28-37]. Actuators of this type commonly utilise advanced materials such as doped-polymers [28], as well as the highly publicised carbon nanotube (CNT) [29, 30] and graphene-based materials [31-37]. These EA materials can exhibit excitation behaviours similar to those of the more macroscopic actuation schemes mentioned earlier, albeit at the micro/nano-scales. Graphene based actuators are the focus of this chapter with other chapters exploring the most recent progress in associated fields.

As a building block for carbon nanotubes, the two-dimensional single atomic carbon sheets of graphene (Fig. 1a) have attracted significant research interest since their discovery in 2004 [38-40]. Owing to its large surface area, low density, high carrier transport mobility, superior mechanical properties, and excellent thermal/chemical stability, graphene is an ideal material for use in mirco/nanoelectromechanical systems (MEMS/NEMS). These are of great interest both for fundamental studies of mechanics at the nanoscale and for a variety of applications, including force [41], position [42], and mass sensing [43]. In addition, the surface-dominated characteristics of graphene provide a unique opportunity to control its bulk physical properties through surface modification, *e.g.*, adsorption, chemical functionalization, and so on. An example outcome may be the fabrication of graphene based multi-functional devices that combine charge and mass sensing within a single molecule [44].

Graphene is also an exceptional nanoscale building block for constructing macroscopic three-dimensional (3D) bulk assemblies for widespread applications [45]. Utilising a colloidal dispersion of graphene oxide sheet forced as a flow through a membrane filter (vacuum filtration technique), Dikin *et al.* fabricated a freestanding graphene oxide paper (Fig. 1b) with a thickness of 1-30μm [46]. The cross-section SEM image revealed well-packed layers across the whole paper thickness. This material has a better stiffness and strength than many other paper-like structures. Applying a similar filtration technique for chemically prepared graphene dispersion, Chen *et al.* successfully produced a mechanically strong, electrically conductive, and biocompatible graphene paper (Fig. 1c) with a similar microstructure as observed in graphene oxide paper [47]. The existence of a corrugated and layered structure [48, 49] in contrast to one that is flat is of interest as the structural corrugation is crucial in preventing the restacking of graphene layers during the filtration process. The inability to restack promotes a highly porous structure that is particularly suitable for electromechanical supercapacitors and actuators.

Three-dimensional graphene cellular foam is another type of macroscopic graphene-based material, which is currently under extensive study [45, 50-57]. This type of 3D structure has an extremely low density, superior specific surface area, and excellent mechanical strength and stiffness. Recently, Qiu *et al.* employed a freeze-casting technique to fabricate superelastic graphene foams as shown in Fig. 1d [45]. The obtained cellular foam can quickly recover its original shape after the application of up to 80% compressive load with a speed up to 7,000 mm/min. After being compressed for 1,000 cycles, no significant structural degradation was observed. This excellent mechanical resilience and fast response rate are highly desired in electromechanical actuators.

The different forms of graphene-based materials offer a significant repertoire in the design of electromechanical actuators that harness the excellent properties of graphene. Recent experiments have seen the use of monolayer graphene, graphene papers, and graphene foams in different actuator designs, showing excellent characteristics and rich novel physics [31-37, 58, 59]. Section 2 of this chapter will provide a brief review on recent experimental progress. Note that this chapter mainly focuses on actuators in which graphene serves as the active components. Graphene-based composites and related actuators are not covered as graphene often serves as a scaffold, thermal, or electrical conductor in these designs [60].

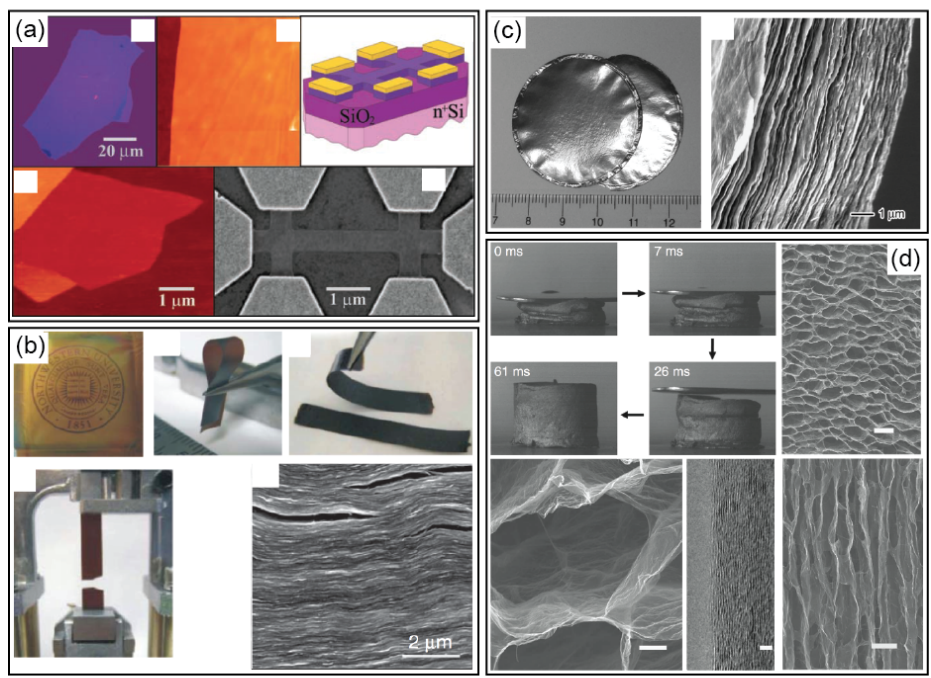

**Figure 1** Graphene and some typical graphene-based materials. (a) Single layer of graphene on SiO$_2$ substrate via mechanical exfoliation from bulk graphite. From [38]. Reprinted with permission from AAAS. (b) Graphene oxide paper fabricated using vacuum filtration technique. Reprinted by permission from Macmillan Publishers Ltd: Nature [46], copyright (2007). (c) Graphene paper fabricated from vacuum filtration. From [47]. Reprinted with permission from John Wiley and Sons. (d) Three-dimensional superelastic graphene monolith. Reprinted with permission from Macmillan Publishers Ltd: Nature Communications [45], copyright (2012).

Before reviewing the recent experimental progress, we would like to provide a brief introduction to three actuation mechanisms for graphene-based actuators: ionic intercalation, electrostatic Coulomb effect, and the quantum-mechanical effect. The purpose is to help readers gain a deeper insight of the performance of different actuators in experiments and clearly understand the advantages/disadvantages. For a graphene actuator immersed in electrolyte (Fig. 2), charge injection of graphene layers will promote three different types of mechanical strain. First, the oppositely charged electrolyte ions intercalate between adjacent graphene layers. As a result, the steric effect causes an expansion in directions perpendicular to the basal planes [61]. Note that this intercalation mechanism is also commonly employed in actuators based on conductive polymeric materials. Whilst the intercalation mechanism generally exhibits favourable characteristics such as low operating voltages, moderate output forces, high strains, and low energy consumption in static applications, research to date has shown that device efficiencies are extremely low (~1%). They also have extremely limited cycling life and slow responses times (due to the ionic diffusion process involved) [31]. Second, for a charged graphene electrode immersed in an electrolyte, the oppositely charged counter-ions will accumulate near the graphene surface to maintain the electro-neutrality. Such a screening region is described as the electrical double layer (EDL). The Coulomb electrostatic interaction within the double layer will generate expansive deformation in the basal plane of each graphene layer, as depicted in Fig. 2b. This EDL effect can generate strain on the order of 1% at low operation voltages (~1V) and allows an extensive cycle live [32, 62]. However, it still exhibits a low response speed (owing to the slow ion diffusion process involved). Third, electrons or holes doped in covalently bonded graphene occupy the bonding or anti-bonding states (Fig. 2c) and thus give rise to a C-C bond length change [62-66]. This mechanical deformation is a result of the quantum mechanical (QM) effect. For a pristine graphene lattice, electron injection generates an in-plane expansion. A low-level hole doping leads to a contraction, but with the increase of hole doping, expansive deformation takes place [30, 62]. Since graphene is an excellent electrical conductor, an extremely short response time can be expected. The robust covalent bonds of graphene should allow an extensive actuator life. However, one clear compromise of the quantum mechanical effect for a pristine graphene actuator is the small strain output, ~0.25% [62].

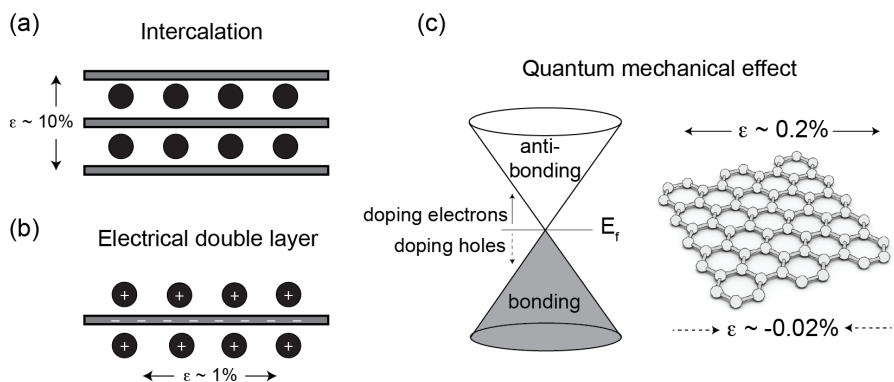

**Figure 2** Actuation mechanisms for graphene and graphene based materials. (a) Electrolyte ion intercalation gives rise to an expansion in a direction perpendicular to the basal planes. (b) The electrical double layer effect causes expansion in the basal plane. (c) The hole or electron injection alters bonding states of the C-C covalent bonds in graphene and thus results in a change of bond length.

## 2. Recent experimental progress

Volume expansion of graphite upon electrolyte ion intercalation is a well-known phenomenon.

Theoretical studies have shown that the insertion of ions among graphite layers can lead to directional large volume expansion (>700%) perpendicular to the basal plane direction of graphite [61]. Utilising this volume expansion for electromechanical actuation has significant potential (Fig. 3a), but is currently limited by fabrication techniques as graphite is commonly in the form of granular particles and is difficult to process into highly ordered macroscopic materials for the design and production of practical devices.

Porous graphene papers assembled from the chemically converted graphene are a promising alternative. This type of paper consists multiple graphene layers that are well aligned (Fig. 3b-c) with a controllable interlayer distance and an in-plane dimension in a macroscopic scale ( > cm). Lu *et al.* successfully demonstrated prototype actuators based on chemically converted graphene membranes using an ionic liquid (BmimBF$_4$) pre-expansion treatment (Fig. 3) [31]. This type of actuator can be activated at low voltages of < 2V, thickness percentage change reaching as high as 98% (Fig 3d). This type of actuators has a very low response rate with optimal performance initiating at very low frequencies, *i.e.*, 0.005Hz to 0.01Hz. Cycling life is also quite low: 600 cycles at 0.05Hz and 30 cycles at 0.01Hz, with an observation that a higher voltage, longer actuation half period, and more cycles accelerated the breakdown of electrodes.

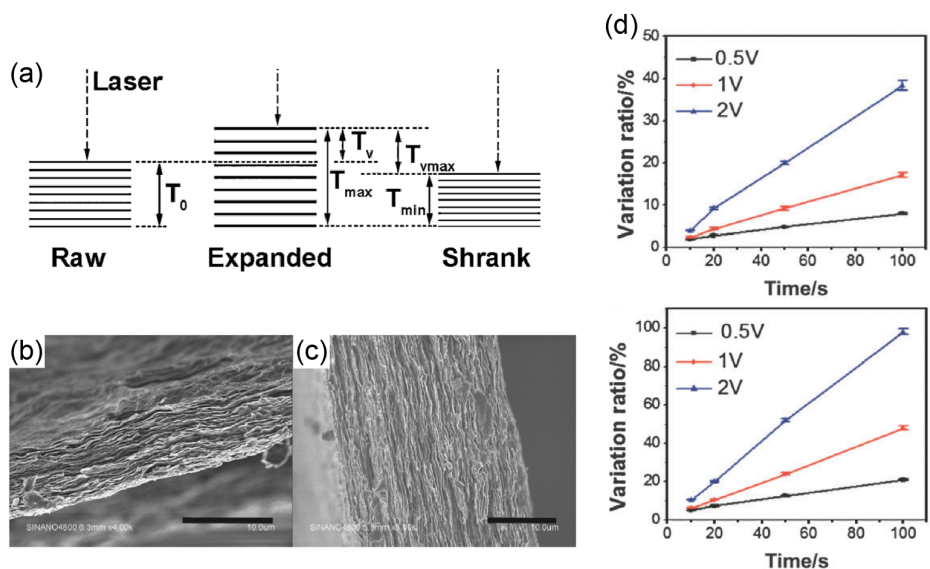

**Figure 3 Electrolyte ion intercalation leading to thickness expansion in graphene membranes.** Adapted from [31] with permission of The Royal Society of Chemistry. (a) Side view sketch of the film thickness change from a raw value T$_0$ to an expanded maximum thickness T$_{max}$ and a shrank minimum thickness T$_{min}$. (b) SEM images for the cross section of a pure RGO film with a scale bar of 10 $\mu$m, (c) SEM image for the cross section

of the RGO/ionic-liquid membrane of 50 wt% ionic liquid with a scale bar of 10 $\mu$m. (d) The thickness change ratio for composite membranes containing 50 wt% or 66.7 wt% ionic liquid, being subject to an AC square electric potential. The applied voltages are 0.5, 1, and 2 V, respectively. The half period varies from 10 to 100 seconds.

Utilizing the in-plane length change of graphene papers to drive actuation is also explored with a focus on unimorph and bimorph designs. Liang *et al.* demonstrated electrochemical actuation for self-supported graphene papers comprised exclusively of graphene flakes [32], whereby the papers (Fig. 4) were fabricated using the filtration process similar to those used in Ref. [46, 47]. The $Fe_3O_4$ nanoparticles were introduced among the graphene layers to prevent restacking of the individual graphene layers and thus enhance the porosity of the graphene papers. Length change of the unimorph graphene paper device immersed in 1M NaCl electrolyte under externally applied cyclic electrical potential (with a magnitude < ±1V) was used to calculate the actuation strain. The graphene papers always elongated under either positive or negative potential, with a parabolic relation between the strain and applied potential revealed (Fig. 4c). A maximal strain of approximately 0.1% is achieved at –1V (Fig. 4c), with strain outputs increasing almost linearly with specific capacitance (Fig. 4d).

These features are consistent with predictions from the Coulomb effect in EDL [62, 67]. Thus the authors concluded that the Coulomb effect was dominant [32]. An interesting feature of the parabolic relationship is an asymmetry between electron and hole injection (Fig. 4c), *i.e.*, higher strains for electron injection. Because the QM strains of graphene change sign with respect to the potential of zero charge, the authors attributed the observed asymmetric strain output to the quantum mechanical effect. This may be an oversimplification. It is known that the graphene cathode (negatively charged) often has a higher capacitance than the anode (positively charged). Such a difference in capacitance could also lead to the asymmetric strain outputs observed in experiments. Despite a relatively thorough understanding of pure graphene paper actuator's performance and function, a comprehensive theoretical model to guide device design is still absent. It is highly desirable to develop a model to predict the electrochemical strain based on the intrinsic mechanical and electrochemical properties of graphene papers.

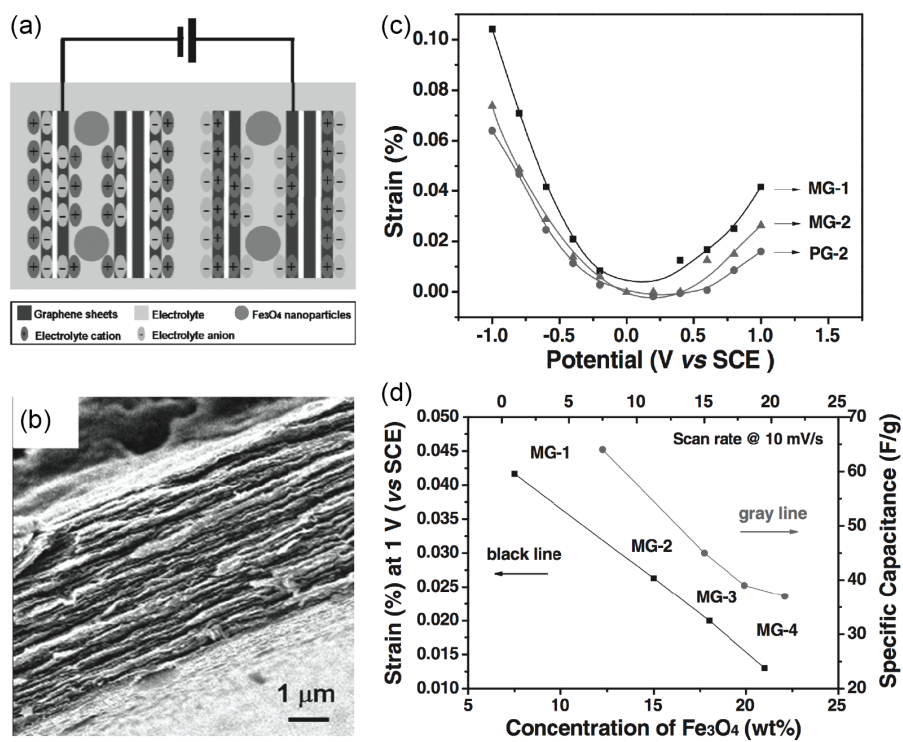

**Figure 4 Graphene paper actuator.** From [32]. Reprinted with permission from John Wiley and Sons. (a-c) Hybrid graphene/Fe$_3$O$_4$-nanoparticle actuator: (a) schematic illustration of charge injection in graphene layers and accumulation of oppositely charged electrolyte ions at the outer surface and inside stacks of graphene. The graphene layers are partially separated by magnetic Fe$_3$O$_4$ nanoparticles. (b) SEM image of the cross-section of graphene paper. (c) Actuation strain of graphene paper as a function of applied voltage in 1 M NaCl solution. (d) Correlation between strain output and specific capacitance of graphene paper actuators.

Graphene is a surface-dominated bulk material with ideally 100% of the atoms on the surface. This provides a unique opportunity to control its bulk physical properties by surface chemical modification. Indeed Xie *et al.* found that after using oxygen and hexane plasma to treat different sides of a graphene paper (Fig. 5a-b), the obtained bimorph actuator (immersed in 1M NaClO$_4$ electrolyte) exhibited a significant bending deformation when subject to electrical potentials as described in Fig. 5c [33]. A maximum 0.2% strain is obtained at 1.2V. Note that their pristine graphene paper actuators showed negligible strain. Through comparative studies of two unimorph devices, in which two sides of each of the graphene papers were treated using only either oxygen or hexane plasma, the authors concluded that the observed actuation of the bimorph and unimorph device solely came from the oxygen treated side. In a following study [34], through optimizing their fabrication process, the same research group obtained a strain output as high as 0.43%. In addition to the larger strain output, the chemically treated graphene papers exhibited two distinctive features from their pristine graphene papers counterparts [32] *i.e.*, (1) elongation upon electron injection whereas contraction being due to hole injection (nearly symmetric about the zero potential as seen in Fig. 5c); (2) a linear relation between strain and applied electrical potential (Fig. 5e). Inspired by feature (1), the authors rationally developed a bimorph PPy/graphene actuator [35]. As seen in Fig. 5f, the PPy displays an opposing strain compared with the graphene layer ($O_2$ plasma treated) when subject to either a positive or negative potential. In this way the synergy of opposing PPy/graphene strains contributed to increased bimorph actuator deformation (Fig. 5g-h). Using a theoretical model, the authors predicted 5 times more strain output assuming that the synergetic effect could be completely achieved in experiment [35]. Additionally, the bimorph PPy/graphene has a 10 times higher actuation rate than graphene papers, with a long actuation life; exhibiting with no significant performance change after 500 cycles.

The distinctive actuation performances between plasma treated graphene paper [33, 34] and pristine graphene paper counterpart [32, 33] indicate that $O_2$ plasma treated surface layers are the main contributor to actuation. The obtained linear strain-voltage relation (Fig. 5e) rules out the Coulomb EDL effect (which should induce a parabolic strain-voltage relation) [62, 67]. The ion intercalation could generate some deformation but mainly in the thickness direction. In light of the qualitative agreement between feature (1) and the trend of QM strain in pristine graphene (section 1), Xie *et al*. attributed the significantly enhanced strains to the quantum mechanical effect. An in-depth analysis, however, suggests possible novel physics for electrochemical actuations in this system. In Xie's experiments, the maximum injected charge was estimated as $-0.01e$/C-atom. According to *ab initio* simulations, the quantum mechanical strain for a pristine graphene is around 0.05% at this charge level [62-64]. Analyzing experimental results of graphene paper actuators also suggests a quantum mechanical strain < 0.03% at a similar charge level [32]. Both estimates are significantly lower than the measured strain of 0.2 – 0.43%. In addition, considering that the internal graphene layers of the paper actuator in Xie's experiments had almost zero strain output, the actuation strain from the $O_2$ plasma treated surface layers should be much higher than the observed strain for the whole paper actuator (*i.e.*, 0.2 – 0.43%). Clearly the QM strain from pristine graphene cannot explain the actuation performance of $O_2$ plasma treated graphene paper actuators. We believe that the superior strain output is rooted in changes to surface molecular structure caused by the $O_2$ plasma treatment. Our recent *ab initio* simulations for graphene oxide showed that epoxy functional groups could significantly enhance the QM strain output (details in section 4) [68]. It is reasonable to expect that $O_2$ plasma can introduce some surface epoxy groups, which contributes to the superior actuation performance displayed in Xie's experiments [33]. Continued work is certainly required to further explore the application of surface chemical modification and help develop high performance graphene based actuator designs.

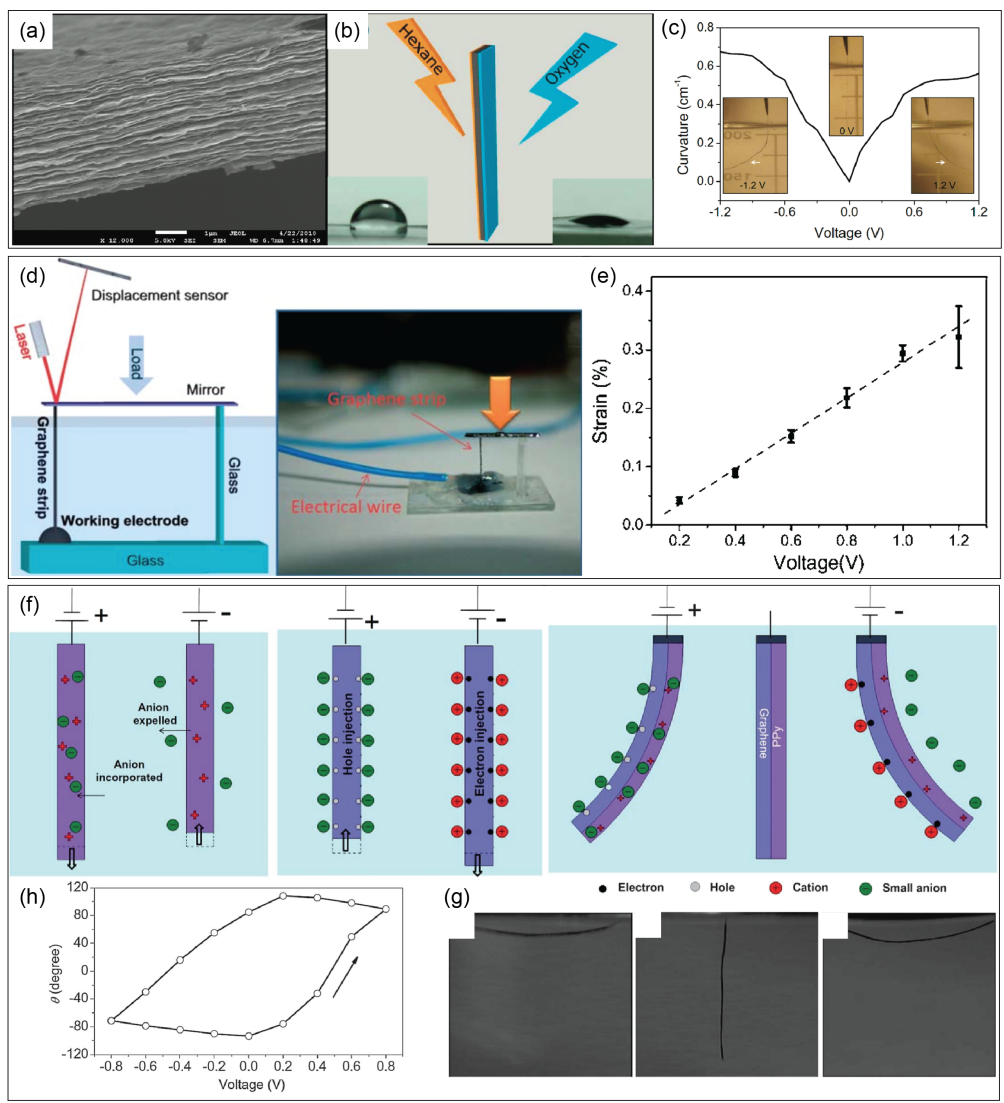

**Figure 5 Chemical functionalization of graphene surface can significantly enhance actuation strain**. (a)-(c) An asymmetric surface functionalized graphene paper actuator. Adapted with permission from [33]. Copyright (2010) American Chemical Society. (a) Cross-section SEM image of a graphene film (scale bar: 1 μm); (b) schematic illustration of asymmetric plasma treatments of the graphene film with hexane and oxygen plasma. The resultant surface wettability is also shown. (c) Curvature change of an asym-modified graphene strip as a function of applied DC potential within 1.2 V. The insets show the status of the graphene strip at −1.2, 0, and 1.2 V, respectively. (d-e) Graphene paper unimorph actuator with $O_2$ plasma treated surfaces. Adapted from [34] with permission of The Royal Society of Chemistry. (d) The apparatus used to characterize the actuation performance of graphene films, which consists of one strip of graphene film fixed on a substrate and connected

to the working electrode. (e) The corresponding strain changes as a function of applied square wave voltages with a frequency of 0.05 Hz. (f)-(h) Synergetic graphene/PPy bimorph actuator. Adapted from [35] with permission of The Royal Society of Chemistry. (f) A schematic illustration of the expansion–contraction mechanisms of PPy and graphene film, and the actuation mechanism of PPy/graphene bimorph structure. (g) Snapshots of the bending deformation for PPy/graphene bimorph actuator driven by electrochemical potential within ±0.8V. (h) The change of $\theta$ angle in response to the applied potential cycle at a scan rate of 50 mV/s in 1 M $NaClO_4$ aqueous solution.

Three-dimensional graphene cellular foam is another promising candidate for electromechanical actuators. Biener *et al.* developed an inherently inexpensive, scalable approach to fabricate a mechanically robust, centimeter sized graphene foams [36]. After utilising sol-gel chemistry to fabricate highly cross-linked organic gels, they transformed the gels to porous $sp^2$ bonded carbon networks through a pyrolysis process. Through a controlled burn-off in an oxidized atmosphere, the more reactive amorphous carbon was removed from the carbon networks with the resultant sample constructed almost entirely from interconnected networks of single layer graphene (Fig. 6a). A macroscopic strain response was observed when applying electrical potential to foam immersed in a 0.1M $NaClO_4$ electrolyte. As the potential cycled between −1V and +1V, amplitude of the strain is 2.2% (Fig. 6b), with a nearly linear relationship between electrical potential and strain. It was found that negative charge produced expansion whereas positive charge induced contraction. In a follow-up study from Shao *et al.* in the same group [69], it was indicated that at a low scan rate (< 1mV/s) the strain amplitude drifts, attributing this irreversible change to the plastic deformation of the graphene foams. When actuated at a higher scan rate (5mV/s), a fully repeatable strain cycle was observed with no obvious strain drift. Further interrogation reveals that the contraction under +1V is significantly smaller than the expansion under −1V, ~0.06% *vs*. ~0.90%, in contrast to the approximately equal magnitude of strain under positive or negative potentials observed in the plasma treated and pristine graphene papers [32, 33]. The characteristic time constant for strain response of about 165s reveals a slow response rate, with the significant drift experienced reducing the actuation life.

From experimental investigation, the strain magnitude of the graphite foam is proportional to the surface area [69]. As the electrolyte ions were intercalated in bulk, intercalation mechanism should not scale with surface area. Subsequently ruled out the intercalation mechanism, the authors then correlated the QM effect with actuation, believing this to be the primary physical mechanism [69]. However, several inconsistencies should be noted. First, the measured strain, 1.1% at about 0.01e/C-atom, is much higher than the pristine graphene paper experiment (~0.03%) [32] and the theoretical *ab initio* simulation predictions (~0.05%) [62-64]. Second, the *ab initio* simulations for pristine graphene predict a nearly equivalent magnitude of strain output at equal positive/negative potentials for low levels of charge injection [62-64], which contradict the experimental results. Third, the observed slow response rate is inconsistent with the intrinsically fast quantum mechanical actuation. Given the impressive features of this carbon foam actuator, further study is required to better understand the underlying actuation physics.

Liu *et al.* presented a durable actuator based on a three-dimensional graphene-polypyrrole (PPy) hybrid foam [37]. During experiments, the maximum magnitude of the cyclic strain reached 2.5% when subjected to a periodic square potential wave of ±0.8V. Volumetric expansion was observed at electron injection and contraction occurred upon a hole injection, with the strain output varying linearly with respect to the applied voltage. This graphene-PPy hybrid foam displays excellent durability. After 3 days operation, the strain output reduced from 0.65% to 0.52%. After 20,000 cycles (11 days), the three-dimensional G-PPy hybrid actuator, displaying no noticeable shape change, was able to produce relatively high strain, 0.45%. The actuation response is however quite slow, exhibiting comparable speeds to the 3D graphene foam discussed above.

In a comparative study, Liu *et al.* found that the pure graphene foam, operated under similar conditions as previously, produced a strain output of 0.15% and pure PPy film (note: currently impossible to fabricated pure 3D PPy foam) exhibited a much large strain 0.45% [37]. Thus it was concluded that the actuation mechanism of a 3D G–PPy actuator was mainly due on the volume change resulting from an electrochemical alternation in oxidation/reduction states of PPy [25]. Supported by the graphene foam framework, PPy layers can cause the volume expansion when small mobile ions are intercalated under an applied positive potential. By contrast, the 3D G–PPy structure will shrink as a result of the expulsion of anions from the PPy body at negative potential. This mechanism goes partway to explaining the observed low response rate.

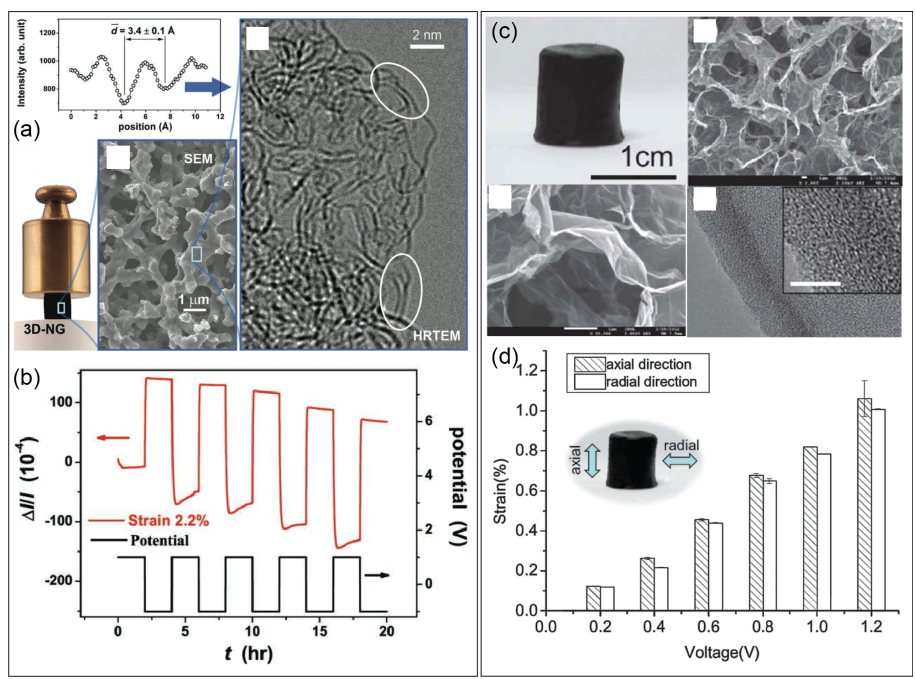

**Figure 6 Actuation of three-dimensional (3D) graphene based foams**. (a-b) 3D nano-graphene based actuator. From [36]. Reprinted with permission from John Wiley and Sons. (a) Illustration of robustness of a 3D nanographene sample, which has an open macroporous network architecture composed of curved and intertwined graphene sheets. (b) Strain and applied electrical potential versus time *t* for the 3D foam immersed in a 0.7M NaF electrolyte solution. (c-d) 3D graphene–polypyrrole hybrid electrochemical actuator. Adapted from [37] with permission of The Royal Society of Chemistry. (c) Digital, SEM, and TEM images of the 3D graphene-polypyrrole porous network. (d) Comparison of the strain response in axial and radial directions of the as-prepared 3D G–PPy as a function of applied square wave voltages of 0.8 V with the cycle period of 50 s.

Lastly, we will briefly review the application of single/few-layer graphene in MEMS/NEMS devices. Shortly after the discovery of graphene, its potential for usage in MEMS/NEMS became a subject of immense research. An initial use by Bunch *et al.* saw mechanically exfoliated multi-layered graphene from graphite deposited over a pre-defined trench on $SiO_2$ substrates [58]. A time-varying radio frequency voltage superposed on top of a static gate voltage was used to initiate vibration of the micron-scale beam/cantilever clamped to the $SiO_2$ surface. The resonance frequency is on the MHz range. The quality factor significantly improves with reduced temperature, ranging from 100 at 300K to 1800 at 50K. This system has superior sensitivity for force and mass sensing. More importantly, they found that the inherent system tension (strain of $2.2 \times 10^{-5}$) in the graphene significantly enhanced the resonance frequency, *e.g.*, from 5.4 to 70.5MHz. The ramifications for this observation suggest a possibility for resonator design with an excellent frequency tunability.

In 2009, Chen *et al.* successfully fabricated monolayer graphene resonators with micrometer sized width and length [44]. A striking feature is that the monolayer graphene resonators possess a unique combination of high resonance frequency and extensive tunability using an electrostatic gate. Tuning the gate voltage can control the tension in the graphene layers and thus leads to a significant change in resonance frequency. It is a unique feature for devices with thickness near the atomic scale such as graphene and single-walled carbon nanotubes, not observed in multi-layer graphene and other NEMS resonators fabricated using top-down deposition/etching techniques. Using a classic model for membrane resonator, and in light of graphene's ability to sustain mechanical strain up to 25-30%, Chen *et al.* proposed that introducing an additional 1% strain in a graphene resonator could allow a GHz frequency range to be achieved. For other NEMS, the enhancement of frequency is often achieved through a reduction in dimension. This is done, however, at the cost of both output signal magnitude and signal amplitude at the onset of nonlinearity, decreasing the dynamic range and creating difficulties with gigahertz-range. The authors estimated the sensitively of their best device sample could approach 2zg.

Suspending graphene layers over trenches requires complex fabrication techniques, which severely limits a large-scale production of resonator devices. Recently van der Zande *et al.* developed a new method to transfer graphene grown by chemical vapor deposition to various substrates [59]. Four different types of resonators were demonstrated. For each type, they were able to produce devices with similar dimensions in a single run, varying from hundreds to hundred of thousands units, with typical resonance frequency was in the MHz range. Consistent results were obtained among similar device samples, indicating a high quality fabrication process. Good frequency tunability with a static gating voltage was also observed. This wafer-scale processing technique provides an encouraging step forwards the practical application of graphene-based devices.

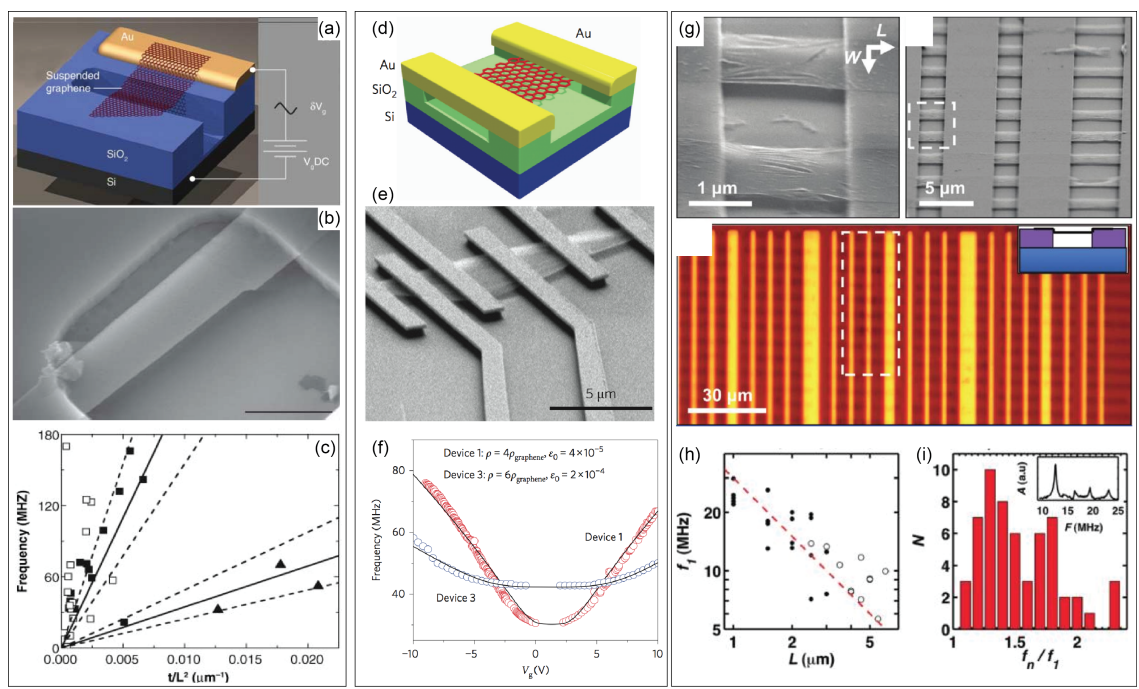

**Figure 7 Resonators based on single or few graphene layers.** (a-c) Graphene sheet electromechanical resonators. From [58]. Reprinted with permission from AAAS. (a) Schematic demonstration of a suspended graphene resonator over a $SiO_2$ trench. (b) Scanning electron microscope image of a resonator composed of few-layer (~2) graphene (scale bar 1 μm). (c) A plot showing the resonance frequency of the fundamental mode of all doubly clamped beams and cantilevers versus $t/L^2$ where $t$ is thickness and $L$ is length. (d-f) Monolayer graphene oscillators. Reprinted by permission from Macmillan Publishers Ltd: Nature Natotechnology [44], copyright (2009). (d) Sketch of suspended graphene over a $SiO_2$ substrate. The $SiO_2$ beneath the graphene flake was removed away through etching. (e) SEM image of the resonator. (f) Measured resonant frequency versus gating voltage $V_g$ for two devices. (g-i) Large scale array of single layer graphene resonators. Adapted with permission from [59]. Copyright (2010) American Chemical Society. (g) Scanning electron microscopy (SEM) image and optical images of suspended graphene membranes over trenches in silicon oxide. (h) Resonance frequency versus length for graphene membrane resonators with widths W between 2.5 and 5 μm. Solid dots are damage-free membranes and open circles are those with partial tears. The dashed line represents the scaling relation $1/L$. (i) Histogram of all measured higher mode frequencies divided by their associated fundamental mode frequencies for 38 identical resonators. The inset shows one typical resonance spectrum.

## 3. Ab initio simulations to reveal the origin of pristine graphene electromechanical actuation

To realize the full potential of graphene based electromechanical actuators, an in-depth understanding of the exact mechanism behind their actuation is vital. In their original manuscript,

Baughman *et al*. postulated that the high elongation strains of SWNT sheets measured upon electrolyte immersion and charging were likely due to some combination of two phenomena: QM (Fig. 2c) and the EDL effects (Fig. 2b) [30]. A brief description of these two mechanisms has been provided in the introduction section. Several groups have sought to investigate the strain due to these two phenomena theoretically, using various approaches including density functional theory (DFT) [62-66]. Two of these groups studied the quantum-mechanical actuation of graphene due to simple charge injection using so-called jellium (uniform compensating background charge) as shown in Fig. 8a [63-65], while the other sought to isolate the two phenomena by simulating the electrostatic EDL as a 2D shell of jellium [66]. However, the latter study could not account for the finite size of the ions present in real EDLs, and thus could not accurately consider the electrostatic interaction between adjacent ions.

Via *ab initio* density functional methods, this section will reveal the physics behind the actuation of monolayer graphene immersed in an ionic liquid (IL) electrolyte, which incorporates the complete ion-ion, ion-electron, and electron-electron interactions that exist in real electrochemical EDLs. Interest in ILs has been intense in recent times, predominantly due to their high energy storage capacities in EDL supercapacitors [70-72]. Their solvent-free nature means that IL-based EDL devices are not affected by solvent presence, unlike aqueous electrolytes, allowing these devices to reach maximum EDL charge concentrations [71]. In addition, ILs show stability across higher electrochemical windows (~5 V) than aqueous and organic electrolytes [71, 72]. As such, it is both timely and topical to investigate the use of ILs in electrochemical actuators. This full ionic and electronic study facilitates the direct probing of the quantum-mechanical and Coulomb EDL effects, and their relative contributions to the overall actuation of covalent carbon materials.

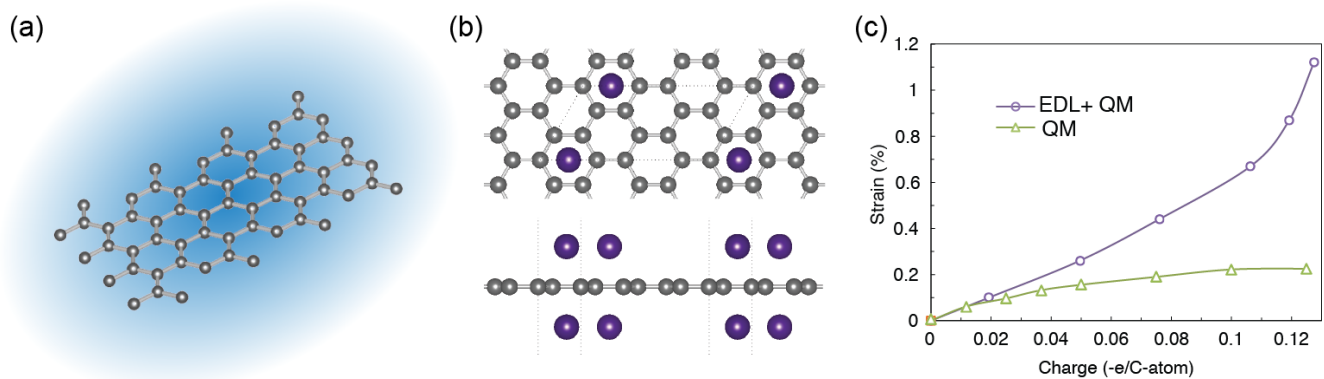

**Figure 8 Atomistic models to study the quantum mechanical and electrical double layer actuation of pristine graphene.** (a) A charged graphene that is embedded in a jellium background with equivalent opposite

charge. (b) A $C_{12}$ unit cell for double layer charging simulation where the large purple spheres represent the ions in double layer. (c) Strains from EDL and QM effects as a function of electron charge injected in graphene. Adapted with permission from [62]. Copyright (2011) American Chemical Society.

DFT calculations were performed using the Vienna ab initio simulation package (VASP v.5.2.2), making use of the projector augmented wave pseudopotentials and the generalized gradient approximation [73-75]. A plane wave cutoff energy of 400 eV was used throughout. Figure 8a-b show the atomistic models used in our *ab initio* simulations. Figure 8a shows a charged graphene embedded in an oppositely charged jellium background. Simulations on this model will examine the QM actuation effect. In Fig. 8b, the EDL ions are located on opposing sides of the graphene monolayer over a C-ring center to simulate the EDL effect. Simulated-annealing molecular dynamic simulations revealed this configuration to be the absolute ground state. Lithium (Li), calcium (Ca), barium (Ba), potassium (K), and sodium (Na) EDL ions were tested. For both models, a unit cell including 12 carbon atoms were used. Note that both QM and EDL effects will contribute to the electromechanical strain output of this unit cell. An 18×36×1 Monkhorst-Pack gamma-centered k-point grid was adopted for all simulations.

To simulate true monolayer graphene using the plane wave code, very high vacuum volumes were included adjacent to the graphene layer to mitigate interlayer electrostatic interactions. Earlier DFT studies employing periodic boundary conditions only separated adjacent graphene layers and individual SWNTs by less than 20 Å [63-65]. Sun *et al*. showed that there exists a considerable amount of variation in the graphene strain per unit charge injected for varying interlayer spacings of up to 12 Å. Our investigations revealed that it is necessary to separate adjacent layers by at least 60 Å in order to simulate isolated monolayer graphene with jellium charge compensation. This is based on the convergence of the graphene electromechanical strain as a function of the interlayer spacing for a given charge injection. Lower values than this are feasible for models incorporating EDLs, such as that of Pastewka *et al.*, as the EDL screens the interlayer electrostatic interactions [66]. As such, an interlayer spacing of 60 Å was adopted throughout for monolayer graphene, providing a good balance between simulation accuracy and computational effort.

To study the EDL charging of a single graphene layer, dopant atoms were introduced aside the graphene in accordance with Fig. 8b to simulate the complete immersion of the graphene in a molten salt IL electrolyte. We employed a molten salt electrolyte rather than a room temperature IL (RTIL) in our DFT calculations in order to circumvent the severe computational expense associated with modeling the latter. Charge injection to the graphene layer was achieved by allowing the electrons to transfer freely between the EDL ions and graphene during self-consistent electronic relaxation. This was done for a range of graphene-EDL ion separation distances (Fig. 8b) in order to simulate varying EDL strengths. In all cases, the graphene and EDL ions were allowed to move freely parallel to the plane of the graphene, but were locationally fixed along the perpendicular. The VASP source code was modified to achieve this, allowing the cell geometry and ionic positions to completely relax in the plane of the graphene layer only.

Given this method of graphene charge injection, it was necessary to establish a procedure to quantitatively measure the charge transferred between the graphene layer and the EDL ions. There are a vast number of approaches available to do this, with much scrutiny as to their accuracy for various systems. For a summary and analysis of the available methods, see the recent article by Manz and Sholl [62, 76]. On the basis of recent demonstrations and reliability [77], we have adopted the Bader method, which calculates the charge density zero flux surfaces and uses these to assign the electronic charges to atoms. This so-called basis set method has been widely applied to many systems over time, producing reliable results on most occasions.

The charge-strain relationships for monolayer graphene in the presence of a simulated Li ion EDL using the Bader method are shown in Fig. 8c. The Ca, Ba, K, and Na EDL ions were tested along with Li, for which the results were found to be very similar. By comparison with earlier computational studies, which predicted graphene strains of 0.3-0.6% for charges of –0.05 e/C-atom [63-65], the present jellium strain (~0.16%) in Fig. 8c appears to be significantly less. However, recall that these previous studies used much smaller vacuum volumes (to separate the graphene layer and its periodic images) in their *ab initio* simulation supercells, which we found result in artificial graphene strain due to interlayer electrostatic interactions and the self-energy contribution of the jellium. Similar jellium strains to these earlier studies have been predicted using the model herein for interlayer spacing of less than 20 Å. As we employ large vacuum volumes adjacent to the graphene layer (60 Å), the present jellium strains more accurately represents true quantum-mechanical actuation resulting from C-C bond expansion due solely to charge injection. In addition to this difference in the magnitude of predicted strain, the present jellium charge-strain relationship differs from those parabolic dependencies reported previously for graphene [65]. Through the inclusion of large vacuum layers adjacent to the graphene layer in the present computational models, interlayer electrostatic Coulomb interactions and jellium self-energy effects are mitigated, which we postulate give rise to the artificial parabolic dependencies predicted by others. This is reinforced by a similar trend described by Sun *et al.* for the jellium charging of graphene, albeit with an interlayer spacing of only 12 Å and higher strain magnitudes [64].

It is difficult to measure the QM strain of graphene in experiments, with currently no direct experimental results available. An estimate from the pristine graphene paper actuator experiments is < 0.03% at a ~0.008e/C-atom electron injection, which agree favorably with our *ab initio* computational results.

Comparing the EDL and QM strain results in Fig. 8c, it is immediately evident that the electrostatic EDL has a significant effect on the graphene strain. Not only does the EDL presence induce higher strains for moderate charge injection (0–0.08 electrons per C-atom), but also it enables the graphene to generate strains in excess of 1%. According to the jellium data, such high strains are not otherwise possible via the quantum-mechanical effect alone. The significance of this result is that it proves that the electrostatic EDL effect is dominant in electrochemical actuators for high charge injection (> –0.08 e/C-atom). While this result is for an IL electrolyte, and thus ultimately represents the optimum EDL strain that could be produced by an electrochemical actuator, we expect that the EDL contribution to the overall strain will still be significant in the case of an aqueous electrolyte. Recent experiments of a graphene paper actuator indeed show a dominant EDL strain in aqueous electrolyte (Fig. 4) [32]. Pastewka *et al.* predicted the strain of a SWNT in the presence of an aqueous electrolyte to be double that in a vacuum for charge injections of greater than –0.1 e/C-atom, which points to a significant contribution to the strain by the aqueous EDL [66]. Baughman *et al.* observed strains of greater than 0.2% for the experimental actuation of aqueous electrolyte filled SWNT sheets, comprising a mechanical entanglement of bundled SWNTs [30]. On the basis of these findings, Baughman *et al.* proceeded to predict strains on the order of 1% for unbundled SWNTs. In the present study, strains of greater than 1% were observed for monolayer graphene in the presence of an IL electrolyte. This quantitatively agrees with the unbundled SWNT strain prediction by Baughman *et al.* since the gravimetric surface area of monolayer graphene is optimal (graphene resembles a perfectly 2D structure), and we employ an IL electrolyte, for which maximum EDL charge concentrations are attainable.

By taking the difference between the strains predicted from *ab initio* simulations for the two models in Fig. 8, it is possible to determine the strain solely caused by the electrical double layer effect. Evidently, the charge-strain and potential-strain relationships are parabolic in Fig. 9, with a steep ascent toward strains in excess of 1%. This can be explained using a simple total energy model, where

$$E_{tot} = \frac{1}{2}ks^2 - q\phi + \frac{1}{2C}q^2 \qquad (1)$$

where *k* is the elastic constant of graphene, *s* is the deformation, *q* is the quantum charge, $\phi$ is the electrostatic potential, and *C* is the differential capacitance. The thermal energy has been assumed negligible, and there is no externally applied force and thus no work term. The internal force (*F*) can be equated to these terms of the energy equation by differentiating it with respect to the deformation (s), giving

$$ks = F + \frac{\alpha}{2}q^2 \qquad (2)$$

where $\alpha$ is a simplified capacitive coupling coefficient. Given that at static equilibrium the internal forces (*F*) reduce to zero, the deformation and thus EDL-induced strain becomes a quadratic function of the injected charge: $s \sim q^2$. This agrees with the *ab initio* simulation results very well (Fig. 9). Such a parabolic relation is consistent to the observation by Liang *et al.* in their graphene papers [32].

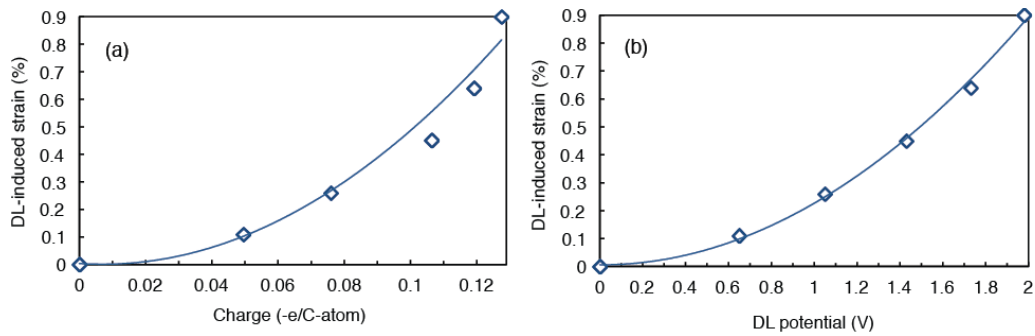

**Figure 9** The EDL-induced strain shown as a function of (a) charge injection, and (b) EDL potential. Adapted with permission from [62]. Copyright (2011) American Chemical Society.

In short summary, *ab initio* simulation and experiment results show that even for moderate graphene charge injection, the contribution of the EDL to the overall strain equaled or exceeded that of the quantum-mechanical strain resulting from charge injection only. The presence of the IL EDL enabled the monolayer graphene to achieve strains in excess of one percent, which was shown to be unobtainable through the quantum- mechanical effect alone. This result may explain the origin of the very high electrochemical strains observed (>0.2%) and predicted (~1%) by others previously for low voltages (<3 V) [30, 78]. To optimize the actuation performance of such materials, it will be imperative to maximize the electrolyte-accessible surface areas to enhance the EDL effect. The use of IL electrolytes in order to reach maximum EDL charge concentrations should prove ideal. A theoretical model based on Eqns. (1) and (2) is under development to predict the EDL strain based on the intrinsic mechanical and electrochemical properties of graphene-based actuators.

## 4. Electromechanical actuations of graphene oxide

The chemical exfoliation of bulk graphite has become a popular method of synthesizing graphene, due to its potential for economical large-scale production [79, 80]. This process involves the oxidation and reduction of crystalline graphite, which leads to the synthesis of graphene oxide (GO) [80, 81] as a prereduction product [79, 80, 82]. GO, due to this ease of bulk manufacture, is readily available and has generated widespread interests for use in many varied applications, including thermomechanical actuation [83]. Studies investigating GO have found that different atomistic structures are attainable, which gives rise to differing electronic and mechanical properties [84-89]. In a recent experimental investigation [85], local GO periodic structures, representative of the highly ordered doping of single oxygen (O) atoms onto the hexagonal lattice of pristine graphene, were observed. Interestingly, approximately 50% of the GO surfaces characterized in this study were found to comprise these novel periodic structures, within which two distinct O atom doping configurations are believed to exist: so-called clamped and unzipped (Fig. 10). In the clamped case, the in-plane lattice constant of the doped graphene was found to be very similar to that of pristine graphene. For this to be possible, it is believed that each dopant O atom binds to two C atoms in the graphene lattice, without rupturing the adjoining C−C bond [85, 86]. For the unzipped case, the in-plane lattice constant is much greater than that of pristine graphene, indicating that the C−C bond is ruptured, unzipping the lattice into conjoined graphene nanoribbons (Fig. 10) [85, 86]. An interesting feature of GO (clamped and unzipped) is that it exhibits a unique structural phenomenon, herein referred to as rippling. As shown in Fig. 10, this prevents the GO lattice from lying completely flat, in contrast to pristine graphene, which relaxes into a nearly perfect 2D plane [86, 88, 90]. The extent of the rippling differs considerably between the clamped and unzipped configurations of GO. We note that the rippling of GO is a short-range periodic effect (∼10 Å), in contrast to the longer range perturbations that are observed in stable monolayer graphene sheets [91].

Given the significant difference in atomistic structure between clamped and unzipped GO, it is foreseeable that each could behave very differently upon electromechanical actuation. For example, if it is possible to modulate the degree of rippling in the respective GO structures, this effect could serve as an origin for a new graphene-based in-plane actuation mechanism. Indeed, our *ab initio* simulations

## 4.1 Electromechanical properties of clamped and unzipped GO

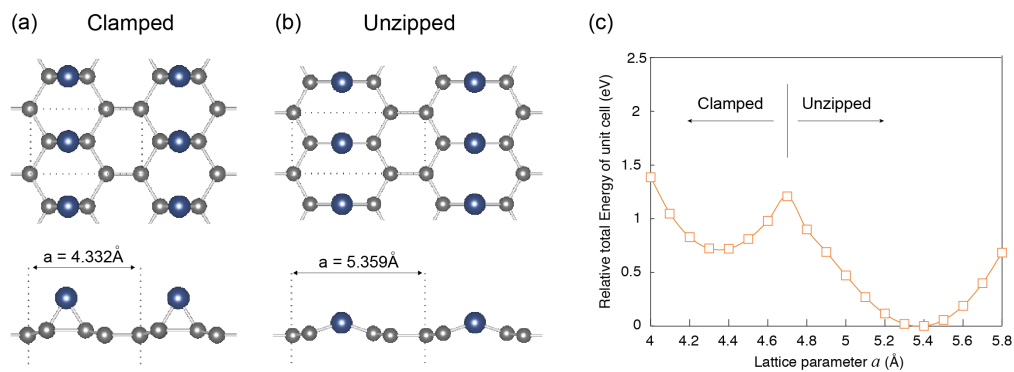

**Figure 10 Symmetrically clamped (a) and unzipped (b) graphene oxide (GO) configurations.** In each case the $C_4O$ unit cells are depicted by dotted lines, with the corresponding in-plane lattice constant *a* shown. The C and O atoms are represented by small silver and large blue spheres, respectively. (c) DFT total energy *vs*. the in plane lattice constant *a*. Adapted with permission from [68]. Copyright (2012) American Chemical Society.

The in-plane strains, measured as the change in lattice parameter *a*, upon charge injection into both GO configurations are shown in Fig. 11. The most pronounced feature is that very high strains are observed for hole (positive charge) injection into GO, especially for the clamped configuration, where a hole injection of 0.15 e/C-atom induces a substantial 28.2% strain. While others have demonstrated intertube electrostatic strains in excess of 10% for strips of aligned CNT sheets [29], the significance of the present 28.2% strain is that it acts along a covalently bonded axis of the material. From Fig. 11 it is evident that the same hole injection into unzipped GO produces strains of 3.6%, which are significantly less than the clamped GO prediction of 28.2%. The origin of this pronounced strain output will be explored later.

Further inspection of Fig. 11 (inset) reveals that electron injection into clamped GO produces a charge−strain relationship very similar to that of pristine graphene (Fig. 8c), exhibiting expansion of up to 0.4% for a −0.15 e/C-atom charge. Conversely, the charge−strain dependency of unzipped GO is distinctly dissimilar to that of clamped GO and pristine graphene, *i.e.*, contraction upon electron injection. This is contrary to what is expected for the quantum-mechanical actuation of covalent carbon materials, such as graphene and CNTs [62-66], where injected electrons are believed to fill antibonding states and thus induce interatomic bond length expansion (Fig. 2c). To explain this peculiar observation, recall that GO has a unique structural property, referred to as rippling, which we hypothesized could give rise to interesting actuation behavior. As this rippling effect is an out-of-plane structural phenomenon, it is possible for the unzipped GO structure to undergo an interatomic bond length expansion, while the unit cell experiences a net contraction along the *a* axis.

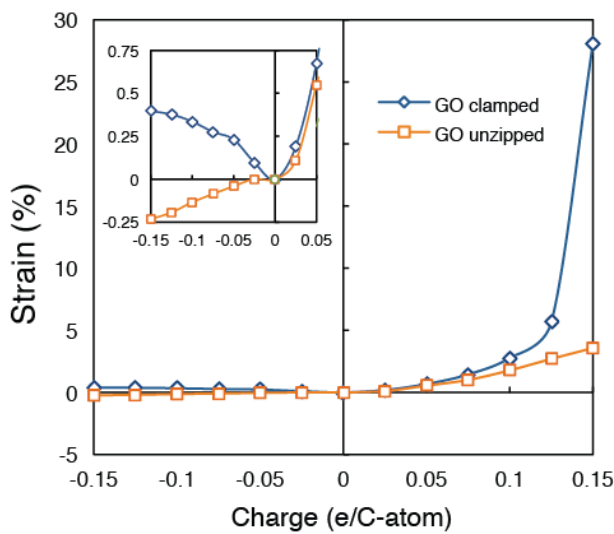

**Figure 11** Actuation of GO (clamped and unzipped) due to positive (hole) and negative (electron) charge injection. Inset: close-up of the strain responses between −0.15 and +0.05 e/C-atom charge injection. Adapted with permission from [68]. Copyright (2012) American Chemical Society.

To investigate, we break down the total strain of the unzipped GO into two contributions: interatomic bond length change and structural rippling change, in Fig. 12b. We quantify the extent of the rippling by the C−O−C bond angle $\alpha$ as indicated in Fig. 12a. For the unzipped GO, an increase in the structural rippling effect, and thus a decrease in the C−O−C bond angle $\alpha$, leads to a contraction of the unit cell along the *a* axis. Here, the interatomic bond length contribution was calculated by summing the individual interatomic bond length changes along the *a* axis for a single unit cell, leading to a prediction of the strain that would be measured if the GO sheet was effectively flat (unrippled). It was then possible to isolate the rippling modulation strain by subtracting the interatomic bond length strains from the total strain. From Fig. 12b, it is clear that rippling modulation has a significant effect on the total strain. The total unzipped GO charge−strain relationship is far more representative of the rippling modulation strain than the interatomic bond length change contribution; at all times having the same strain sign (expansion/contraction). This is also supported by an observed strong positive correlation between the total strain response of unzipped GO and the change in structural rippling, as defined by changes in the C−O−C bond angle (Fig. 12c). This demonstrates that it is possible to generate unique and high strain responses via modulation of the structural rippling for unzipped GO.

To verify this proposed unzipped GO contraction origin, which is quantum-mechanical in nature, it is necessary to consider the excess charge density distributions. In Fig. 12a, the −0.15 e/C-atom injected charge aggregates atop the O atom and also atop and beneath the O-bonded C atoms. It is evident that the excess charge atop the C atoms is repelled by that atop the O atom, as the charge contours lean away from the O atom. This repulsive force between the bonded O and C atoms results in a torque about the O atom, which causes the bond angle $\alpha$ to decrease and the rippling to increase.

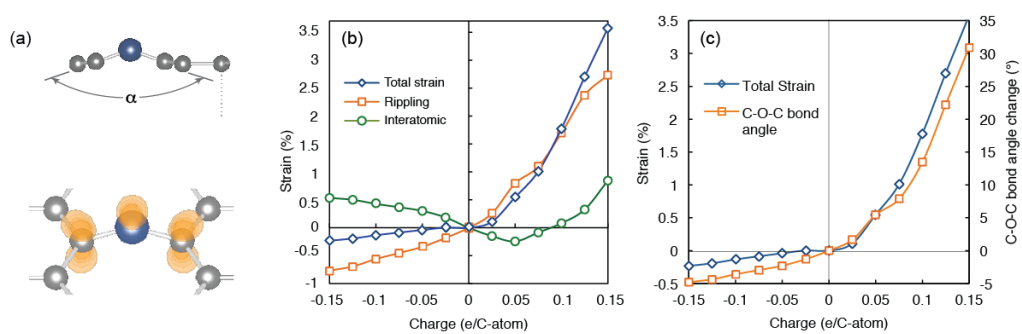

**Figure 12** (a) The excess charge density profiles for unzipped GO explain the observed actuation behavior upon electron injection (orange regions represent excess electron density). Tuning rippling can induce contraction upon electron injection. (b) Total, interatomic and rippling strains of unzipped GO as a function of charge injection. (c) A good correlation between C-O-C bond angle variation (rippling) and total strain output with the change of charge injection. Adapted with permission from [68]. Copyright (2012) American Chemical Society.

To explain the huge 28.2% strain performance of clamped GO, Fig. 13 shows the energy configuration of $C_4O$ GO as a function of the lattice parameter $a$ and the extent of hole injection. As presented and discussed by Xu and Xue [86], the clamped structure represents a so-called metastable phase of GO, while the unzipped case is more stable. Despite the higher stability of the unzipped configuration, it is nonetheless possible to synthesize clamped GO for use in practical MEMS/NEMS actuators [85], due to the relatively high energy barrier (0.63 eV per unit cell) that separates these two states. To understand the origin of the measured 28.2% strain, consider the 0.63 eV per unit cell energy barrier between the clamped and unzipped phases (between $a$ values of 4.33 and 4.7 Å in Fig. 13). Upon hole injection into clamped GO, the energy profile is modified (Fig. 13), resulting in an expansion of the unit cell along the $a$ axis. This expansion is due to the concentration of excess holes on the O atom, as well as at and between the O-bonded C atom sites, which weakens the adjoining C−C bond (Fig. 13a). With further hole injection, the energy profile continues to change until the energy barrier between the clamped and unzipped configurations disappears, and the clamped unit cell snaps from an initial lattice constant of 4.33 to 5.55 Å (while charged). This is the point at which the bond between the O-bonded C atoms is weakened to such an extent that it ruptures, which is the configurational origin of the predicted 28.2% strain. An interesting feature of this actuation mechanism is that it is possible to charge the material causing up to 28.2% strain, and then remove the input power without the material relieving to its original zero strain state ($a$ will relax to 5.36 Å upon charge removal, which corresponds to a 23.8% strain). This GO material will maintain its new unzipped configuration once the C−C bonds beneath the O atoms have been ruptured and the new C−O−C bonds formed. This feature would be particularly beneficial for long-term, low-power switching applications.

Practically speaking, it is desirable that induced actuation be fully reversible. Due to the very high energy barrier transitioning from the unzipped to clamped configurations in Fig. 13 (1.33 eV between $a$ values of 4.7 and 5.36 Å), it would seem difficult to reverse the 28.2% strain. To do so would require an increase in the extent of the unzipped GO rippling and thus a decrease in the C−O−C bond angle $\alpha$, which can only be achieved through the injection of electrons as discussed earlier. From Fig. 13c, the lattice parameter $a$ would need to be reduced by 11.4% from 5.36 to 4.75 Å. It is evident that even the maximum electron injection level (−0.15 e/C-atom) has little effect on the overall energy profile (Fig. 13), which agrees with the results of Fig. 11 (a strain of only −0.23% is attainable for −0.15 e/C-atom electron injection). Hence, reversal of the 28.2% strain would require some additional influence to charge injection alone. Nonetheless, if only irreversible actuation of GO by 28% were attainable, this would still be extremely useful for select applications, such as legislated single-use industrial safety switches. In addition, GO can be used for the generation of high reversible strains in a more traditional actuation setting, with clamped GO capable of generating peak strains of up to 6.3% prior to surpassing its transitional energy barrier and unzipped GO being capable of both large contraction (−0.25%) and expansion (3.6%).

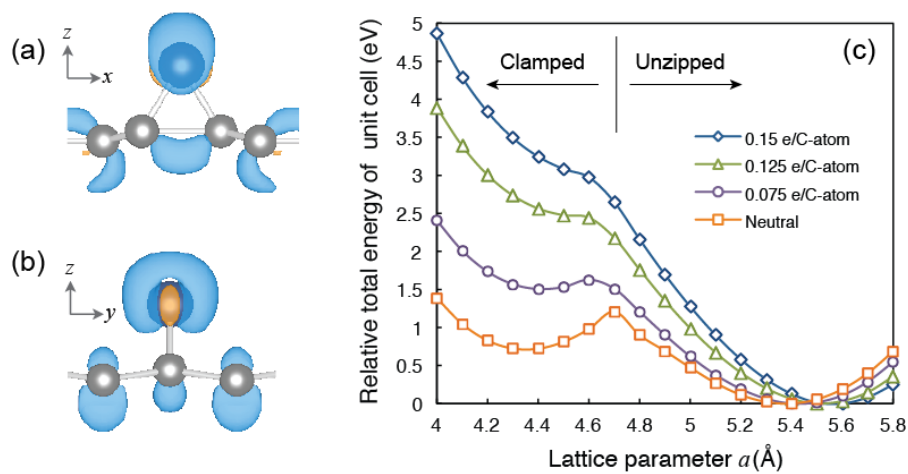

**Figure 13** (a-b) The excess charge density profile of clamped GO upon 0.15 e/C-atom hole injection explains the observed expansion and eventual unzipping of GO (blue (orange) regions represent excess hole (electron) densities): (a) view along the zigzag (*y*) direction; (b) view along the armchair (*x*) direction. Unzipping leads to substantial strain. (c) Relative total energy of the GO unit cell as a function of the in-plane lattice parameter *a* and injected charge. The regions corresponding to the metastable clamped and more stable unzipped configurations are as indicated. Adapted with permission from [68]. Copyright (2012) American Chemical Society.

## 4.2 Graphene oxide molecules as ideal building blocks for artificial muscle

A long-standing challenge in actuation research is the development of an actuation material that mimics the behaviour of mammalian skeletal muscle (a so-called "artificial muscle"). In order for a material to mimic skeletal muscle, it must be capable of delivering sufficient strain (20%–40%) and volumetric work density (0.008–0.04 J/cm$^3$), at reasonably fast rate (>50%/s) [92]. Additionally, a true artificial muscle would not expand upon stimulation as most existing actuation technologies do (elongation expansion actuators are prone to buckling in high load applications), but rather would contract in the same way as skeletal muscle. Currently, scarce few actuation materials that satisfy one or more of these criteria are in existence. The unzipped GO molecule shown in the previous section has demonstrated the appealing features that mimic the natural skeleton muscle, such as contraction upon electron injection and large reversible QM strain (up to 3.6%). However, these are substantially lower than those exhibited by natural muscle. In this section, we will present a special configuration of the unzipped GO that could serve as an ideal building block for artificial muscle. This molecular configuration, through *ab initio* simulations, indicates a reversible contraction strain up to –4.8% and volumetric work density several orders of magnitude higher than natural skeleton muscle.

Figure 14a depicts the molecular structure side-view. In contrast to the unzipped GO shown in Fig. 10 where oxygen atoms are distributed on a single side of the molecule, the oxygen atoms in this structure are distributed on both sides, producing a zigzag profile. Following the terminology used by Xu and Xue [86], we henceforth term this configuration as asymmetric unzipped GO.

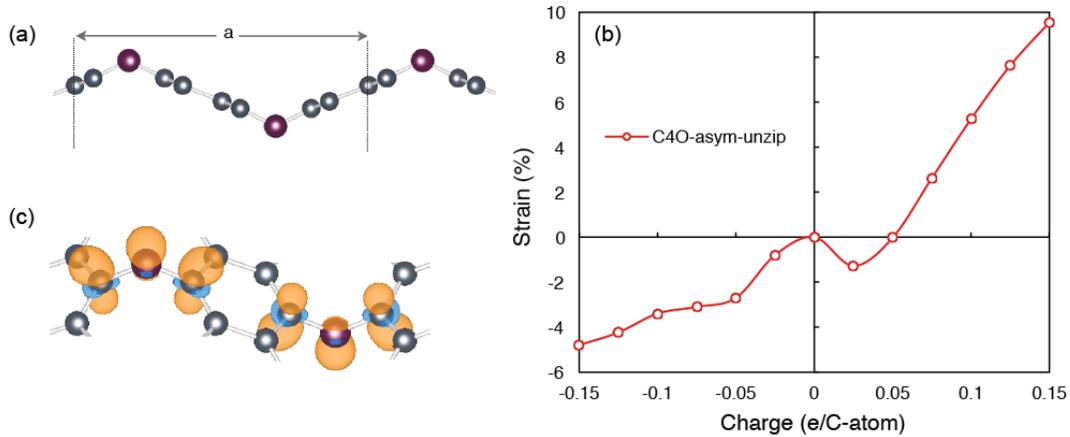

**Figure 14** (a) Side view of $C_4O$-asym-unzip GO. The dashed lines enclose one unit cell. (b) Electromechanical actuation due to positive (hole) and negative (electron) charge injection. (c) The excess charge density distribution of $C_4O$-asym-unzip shows the origin of the very high electron-induced contraction, where orange (blue) regions represent excess electron (hole) concentrations at an electron (hole) injection level of $-0.05$ e/C-atom ($+0.05$ e/C-atom). Reprinted with permission from [97]. Copyright 2013, AIP Publishing LLC.

The electromechanical response of monolayer $C_4O$-asym-unzip GO upon charge injection is shown in Fig. 14b. The first noteworthy result is that $C_4O$-asym-unzip GO exhibits an extremely high charge–strain sensitivity, in contrast to pristine graphene. This particular GO configuration generates a contraction of $-4.8\%$ upon $-0.15$ e/C-atom charge injection; an electron-induced response which is more than five times greater in magnitude than any other GO configuration reported in previous sections. Further insight into the high magnitude contraction of $C_4O$-asym-unzip GO can be garnered by considering the excess charge density distribution upon electron injection in Fig. 14c. Evidently, the excess injected electrons (shown in orange) aggregate atop the O atom, and atop and beneath the O-bonded C atoms. As most of the excess injected charge collects atop these atoms, a large repulsive interaction between these electronic regions exists, which causes the C–O–C bond angle to decrease in order to lower the total energy. This bond angle decrease leads to an increase in the degree of rippling within the GO layer, which is responsible for the shrinking of the unit cell along the $a$-axis. This is quite similar to the observation in Fig. 12. An intriguing observation is that in Fig. 14c there exist regions of excess hole concentrations (shown in blue). This indicates that additional electrons to those injected could in fact be contributing to the overall response, such that there would be two contributions to the total contraction:(1) an "extrinsic" contribution (due to the injection of foreign electrons), and (2) an "intrinsic" contribution (due to the redistribution of domestic valence electrons). Such an intrinsic electromechanical actuation has not been observed in other materials before.

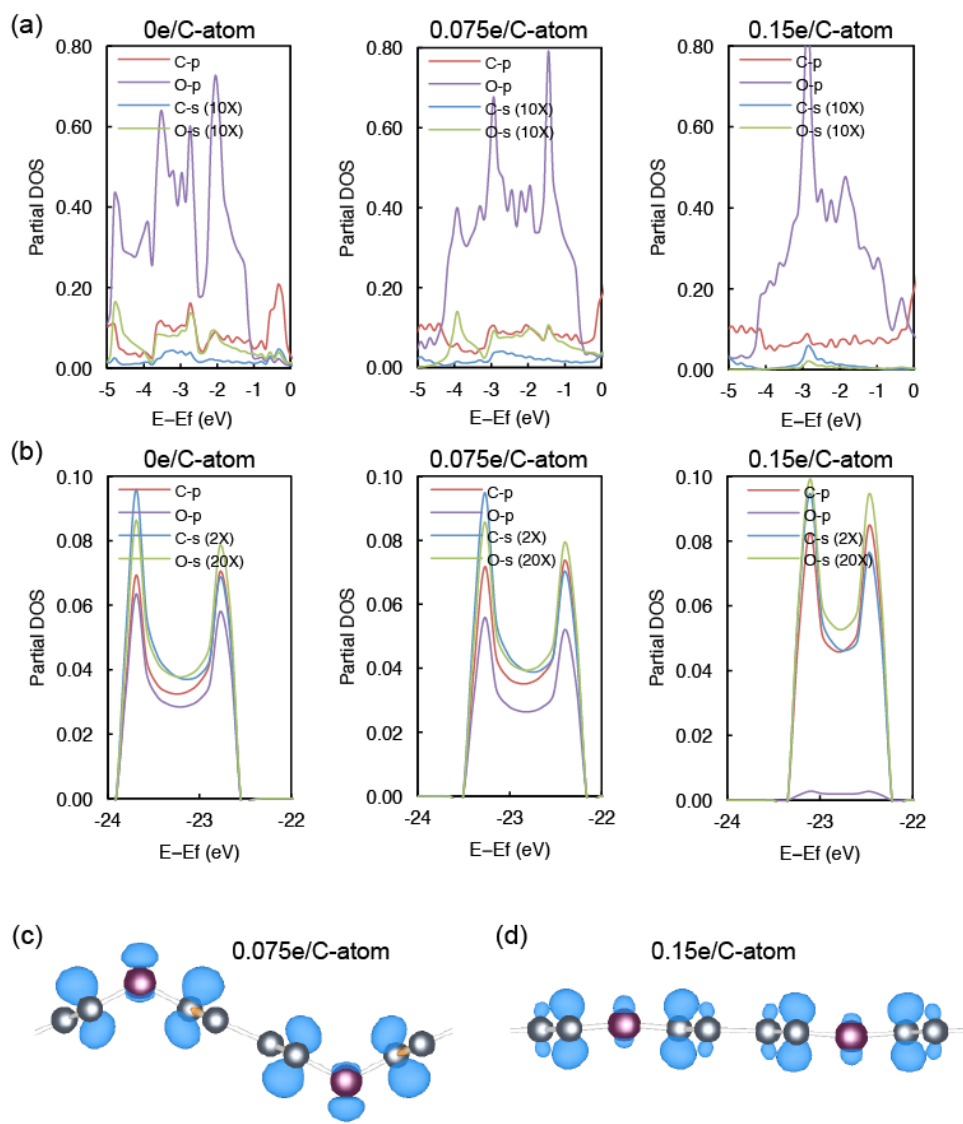

**Figure 15** Partial DOS (a-b) of $C_4O$-asym-unzip GO for 0, 0.075, and 0.15 e/C-atom hole injection. The partial DOS are decomposed into both orbital (s and p) and site (C and O-atom) projections, where the C-atom DOS shown is for the O-atom's nearest neighbour. Excess charge density distribution is also shown for (c) 0.075 and (d) 0.15 e/ C-atom hole injection. Blue (orange) regions signify excess hole (electron) concentrations. The excess

charge density isosurfaces in (c) and (d) are equivalent (normalised against the injected charge). Reprinted with permission from [97]. Copyright 2013, AIP Publishing LLC.

Hole injection into this GO configuration is also rather interesting. Upon low concentration hole injection (0.025 e/C-atom), contraction (−1.3%) is initially observed followed by large expansion (up to 9.6%) with further hole injection (Fig. 14b). The origin of this behavior can be explained by considering the molecular orbitals associated with the C–O–C bond. To this end, we have calculated the spd- and site-projected density of states (DOS) upon hole injection, as well as the corresponding excess charge density distributions shown in Fig. 15. Analogous to the case of the water molecule ($H_2O$) [93], the O atom in $C_4O$- asym-unzip GO has two $sp^2$-like atomic orbitals in the C–O–C plane, which form $\sigma$ bonds with the $sp^2$ atomic orbitals of the C atoms. The O atom has a further two lone electron pairs, one of which exists purely as a p-orbital with its electron density perpendicular to the C–O–C plane, while the other is close to an $sp^2$ orbital in the C–O–C plane. As is evident in Fig. 15a, the O atom has non-zero s- and p-DOS near the Fermi level ($E_f = 0$ eV). This is further confirmed by the excess charge density distributions for low concentration hole injection (Fig. 15c), where there exists a clear $sp^2$-like orbital surrounding the O atom. The C–O $\sigma$ bonds should correspond to energies of between −15 and −5eV (for $E_f = 0$), where the s- and p-DOS of both the C and O atoms overlap significantly. Thus, the two lone pair orbitals of the O atom should correspond to energy levels of −5 to 0eV in Fig. 15a.

With a moderate level of hole injection (*e.g.*, 0.075 e/C- atom), the electron density of the C-atom's p orbital, and the O-atom's $sp^2$ lone pair orbital, shrinks (Figs. 15a and 15c). We believe that such a reduction in the overlap of these two orbitals reduces the repulsive interaction between them, thus, leading to an increase in the C–O–C bond angle and an expansion of the unit cell along the basal plane. For high concentration hole injections (0.15 e/C-atom), significant changes in the DOS are observed. Near the Fermi level, the p-DOS of the O atom significantly increases with a reduction in the s-DOS. Meanwhile, between energy levels of −22.5 and −24 eV, we observe an increase in the s-DOS and a clear reduction in the p-DOS of the O atom (Fig. 15b). This suggests that a hole injection at this level will change the $sp^2$- like nature of the lone pair to a more p-like orbital, as supported by the excess charge density plot in Fig. 15d. In this case, the O atom becomes more C-like in the GO lattice, having both $\sigma$ and $\pi$ bonds with its nearest neighbours. As a result, the GO structure becomes flat.

As for the low concentration hole (0.025e/C-atom) induced contraction of $C_4O$-asym-unzip GO, we believe that the above molecular orbital discussion is, by itself, insufficient to fully describe this phenomenon, which is similar to that seen in pristine graphene (albeit with a much greater magnitude). We find that a quasi-linear relationship exists between the charge injection and the C–O interatomic bond length change, where hole injection always leads to a contraction of the C–O bond. This C–O bond length contraction tends to enhance the overlap of C atom's p orbital with the O atom's $sp^2$ lone pair orbital. As such, we postulate that for low concentration hole injection, this effect overwhelms the shrinkage of the two orbitals due to the removal of electrons, and thus, results in an observable decrease of the C–O–C bond angle.

To assess the potential suitability of $C_4O$-asym-unzip as a building block for the development of artificial muscles, consider the strain ($\varepsilon$), strain-voltage coefficient ($S_v$), and volumetric work density ($W_{vol}$) comparisons in Table 2. Next to mammalian skeletal muscle, $C_4O$-asym-unzip GO produces the highest electromechanical strains. This is especially significant when we consider that muscle operates via contraction, rather than expansion, and this particular GO compound also produces significantly larger contraction values than previously reported. In terms of $S_v$, in order to produce strains of ≥20% at safe voltages of ≤20 V, values of ≥1%/V are required. From Table 2, several materials satisfy this requirement. The same can be said of the $W_{vol}$ values, with a multitude of materials producing higher work densities than skeletal muscle, including $C_4O$-asym-unzip GO (with measured stresses of ~100 GPa). However, the stalling point for all of these alternative artificial muscle candidates is that their producible strains are much too low to be useful, unlike $C_4O$-asym-unzip GO. Whilst the present contraction of –4.8% is lower than mammalian muscle (–20%), we expect that further improvements to this value will follow in the near future. Finally, because the herein predicted actuation responses are QM in origin, we expect that the response times should be on the order of 1 ns, which further cements the potential use of this material as a building block for the development of artificial muscles.

Table 2 Electromechanical strain $\varepsilon$ (%), strain-voltage coefficient $S_v$ (%/V), and volumetric work density $W_{vol}$ (J/cm$^3$) comparison between GO and other materials. Reprinted with permission from [97]. Copyright 2013, AIP Publishing LLC.

|  | $\varepsilon$ (e$^-$/h$^+$) | $S_v$ | $W_{vol}$ (e$^-$/h$^+$) |
|---|---|---|---|
| **Mammalian skeletal muscle** | 20–40 | N/A | 0.008-0.04 |
| **GO-C$_4$O-asym-unzip** | –4.8/9.6 | 7.9 | 4.2/144.1 |
| **GO-C$_4$O-sym-clamp** | 0.4/6.3 | 2.1 | 1.2/52.9 |
| **Pristine graphene** | 0.2/4.7 | 1.6 | 0.3/54.1 |
| **CNT bucky paper** | 0.1 | 0.11 | ~30 |
| **CNT aerogel sheets** | ~1.5 | 0.0015 | 0.00005 |
| **Piezoelectric ceramics** | 0.1 | 0.01 | 0.1-1.0 |
| **Magnetostrictive materials** | 0.2 | ~0.002 | 0.2 |
| **Electrostrictive polymers** | 4.0 | ~1.0 | 0.1-0.3 |

Experimental verification of our *ab initio* simulation predictions is not feasible at present. We require successful fabrication of a sample of graphene with an ordered distribution of oxygen functional groups in a sufficiently large area of its basal plane [85]. This is a challenge for current GO fabrication techniques. As reviewed in section 2 (Fig. 5), oxygen plasma treatment of graphene paper surfaces leads to a significantly enhanced strain output compared with the pristine graphene paper actuators. This is analogous to the trend predicted by our simulations. Since there is no knowledge of the molecular structure after the O$_2$ plasma treatment, no solid conclusions can be drawn.

Our *ab initio* simulations for the ordered clamped and unzipped molecules expose a great potential for graphene-oxide in electromechanical actuation applications. These GO molecules have a much higher actuation strain/stress or work density output than other carbon based materials (Table 2). They also exhibit distinctive features, such as contraction upon electron injection, to address the demands of some special applications, such as its use as an artificial muscle. The second example could be a monolayer graphene-oxide resonator with GHz frequency. Predicted by Chen *et al.* [44], introducing 1% in-built strain in graphene can enhance the resonance frequency from MHz to GHz. The impressive contraction strain –4.8% of the C$_4$O molecule provides a promising route to achieve this goal. The third example is a biomimetic molecular robot. A proof-of-concept design will be presented in next section. It is reasonable to anticipate that the vast diversity of graphene oxide molecular structures will bring a rich material database for micro/nano-actuation materials design that will fulfill requirements for the coming decades.

## 4.3 Design of biomimetic molecular robotic worms

Design of molecular robots at a nanometer scale is an interesting but challenging topic. After millions of years' evolution, strategy adopted by nature can provide valuable inspirations. As shown in Fig. 16a, inchworms exhibit flexible, robust and stable mobility that is far superior to many man-made systems. The inchworm adopts simple but extraordinarily powerful propulsion, with the many small foot pads placed at either end of its elongated body at its disposal (Fig. 16). Thus the inchworm's mode of locomotion is to firmly attach the rear portion of its body to a surface via its foot pads, *extending* the remainder of its body forward, attaching it to the new area of surface and pulling the rear part of its body to meet its forward section through *contraction*. The asym-unzipped GO molecule, exhibiting similar extension/contraction characteristics, could be an ideal candidate to mimic the motion of inchworms (Fig. 16b).

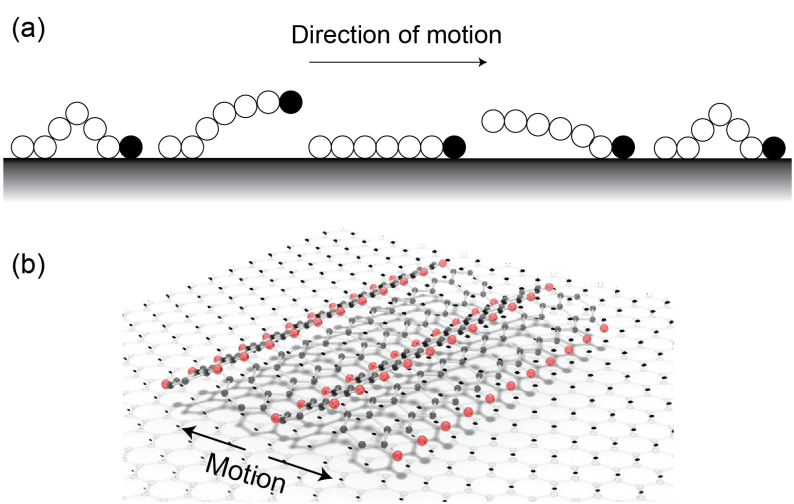

**Figure 16** (a) Sketch of the motion of inchworm after reference [98]. (b) Proof-of-concept design of a GO molecular worm over a graphene substrate. Carbon atoms are shown as black balls with red balls representing oxygen atoms.

In order to achieve unidirectional motion, the foot pads on both head and tail ends must be 'asymmetric'. In other words, during body contraction the head end foot pads grip the surface substrate, whereas during extension the tail end foot pads are engaged in gripping. At all times in motion one end grips the surface, whilst the opposing end foot pads release their hold. In order to mimic this motion, we must realise the asymmetric gripping effects at both ends of a GO molecule (with a finite length).

One appealing solution is to attach different chemical functional groups at either end of GO molecule. Through differing chemical interactions between these chemical functional groups and a substrate, the desired asymmetric gripping effects can be achieved. In this section, we will present a proof-of-concept theoretical study for some prototype designs.

Figure 17 depicts two prototypes of asym-unzipped GO molecular robots. The selected end chemical groups are shown. Since they involve two and four unit cells of the original GO molecule, we term them as two-unit molecule and four-unit molecule, respectively. Longer molecules were also considered for some end chemical groups.

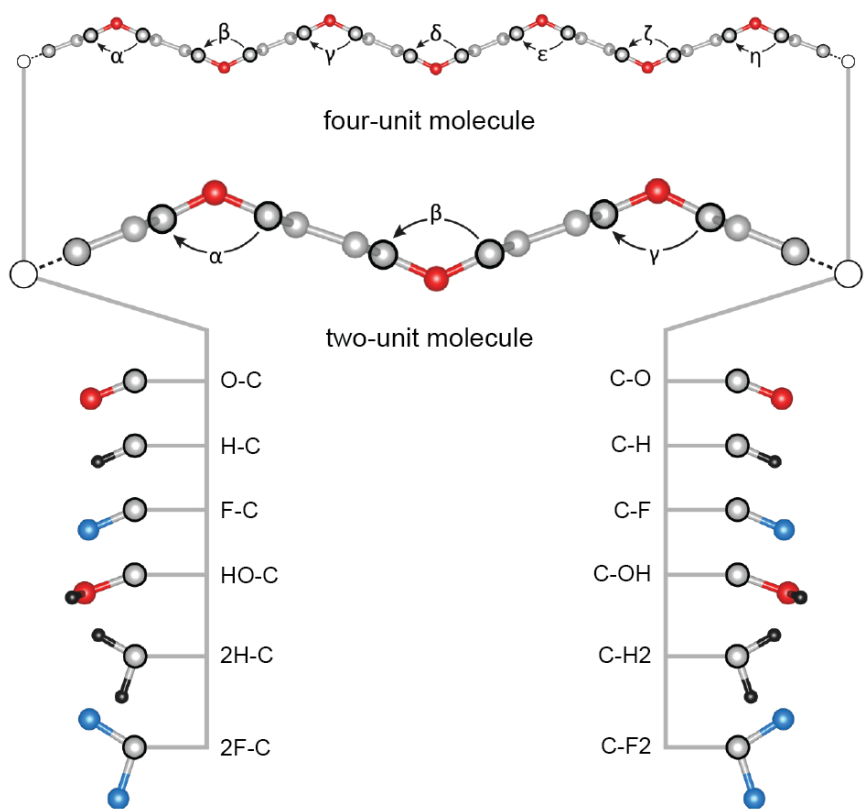

**Figure 17** Prototype GO molecular robots — two asym-unzipped GO molecules with different finite lengths. Both ends are functionalized with different chemical groups.

To investigate the effect of end chemical group substitution on the quantum mechanical strain output, we undertake *ab initio* simulations. Similar simulation parameters were adopted as those used in previous unzipped GO investigations discussed in section 4.1, with all the possible combinations of end chemical groups depicted in Fig. 17 tested. For a charge injection of $-0.15e$/C-atom, results of the contraction strain and the averaged bond angle change are summarized in Fig. 18. The rippling effect consistent with our previous conclusion is further reinforced with these results, exhibiting a strong correlation between the strain and bond angle change. Inspection of Fig. 18 reveals some general patterns. The combinations of –OH and –F groups always yield the highest strain output in both two-unit and four-unit molecules, whereas the combinations of –F2 and –O groups have the lowest strain outputs. It appears that incorporation of either –F2 or –O group leads to a reduction of the electro-chemical strain.

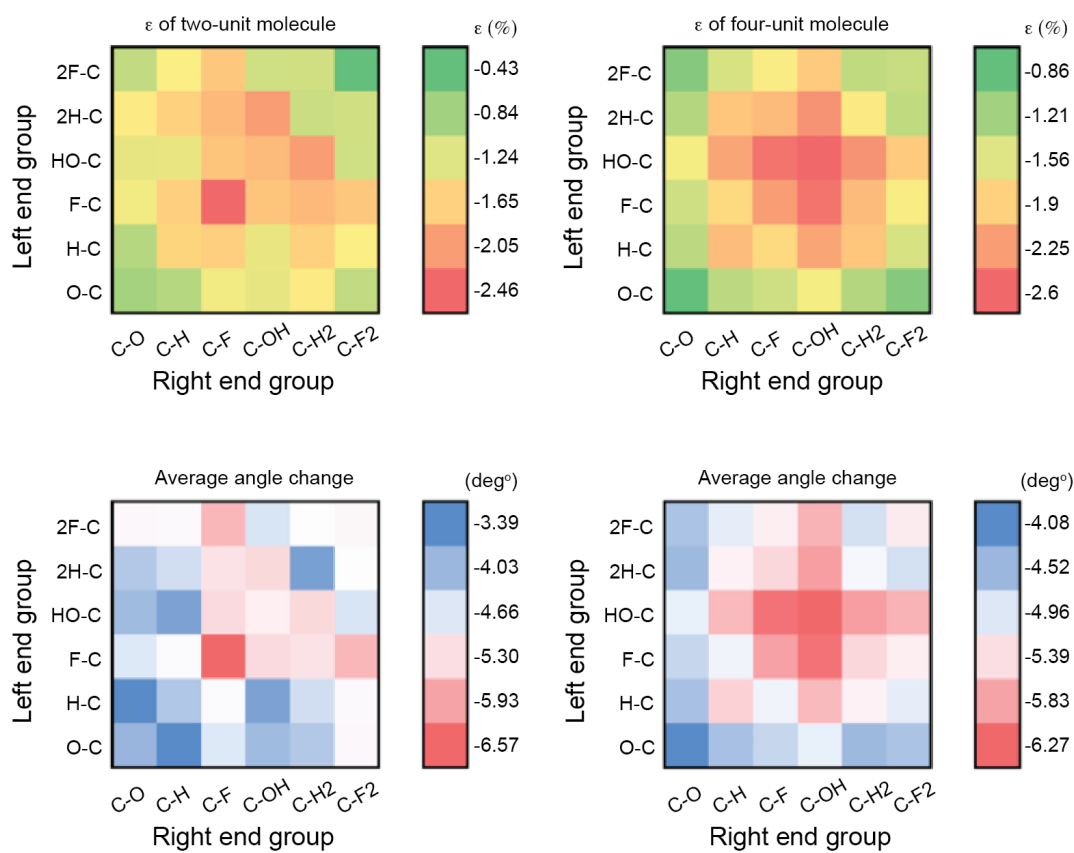

**Figure 18** Electromechanical strain of two-unit and four-unit C$_4$O molecules with different end chemical groups at an electron injection of 0.15e/C-atom.

It is worth noting that the strain output of both GO molecules is significantly lower than that of the infinite-length GO molecule (Fig. 14b). This could be attributed to the edge effects. Taking a two-unit GO molecule with an –OH end group as an example, as shown in Fig. 19a, a greater number of injected excess charges accumulate around the edge –OH group compared with the C-O-C bond. With less accumulated electrons, the contraction of the rippling C-O-C bond angle is significantly lower, leading to a smaller contractile strain output. Figure 19b summarises the dependence of strain on the inverse length of the GO molecules. With an increase in molecular length, the magnitude of strain predictably increases, approaching the result of the asym-unzipped GO molecule with an infinity length.

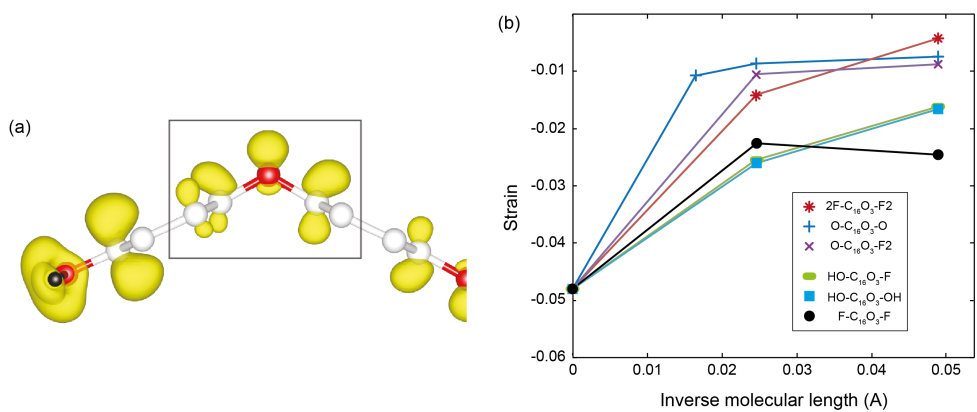

**Figure 19** (a) Excess-electron density for the two-unit $C_4O$ molecule with both ends terminated with hydroxyl groups. (b) The electromechanical strain as a function of inverse length for GO molecules with different end chemical groups.

To simulate inch-worm like motion of the molecules, a two-unit molecule was assembled atop a graphene substrate (Fig. 20). The top three molecules (which give rise to an optimal electromechanical strain output in Fig. 18) with asymmetric end chemical groups were selected, *i.e.*, 2H-$C_{16}O_3$-OH, F-$C_{16}O_3$-OH, and 2H-$C_{16}O_3$-F. The *ab initio* molecular dynamic simulation has the capability to model the dynamic motion of molecular robot, however owing to inherently high computational demands, a fully dynamic investigation was considered unfeasible. Here we assume a quasi-static motion and neglect the inertial effects, with a three-step procedure utilised. First, the molecules were fully relaxed under a charge neutral state. Then, with charge injection at a level of 0.15e/C-atom, the two-unit molecules were allowed to move and find their most stable position. Following the second step, the excess charges were removed and the molecules were able to relax again. Note that the graphene substrate was fixed in our simulations. The coordinate change of the molecular center of mass between step 1 and step 3 was used to represent displacement of the whole molecule in one charge-discharge cycle. The obtained semi-quantitative estimate of the molecule motion verifies our conceptual prototype design.

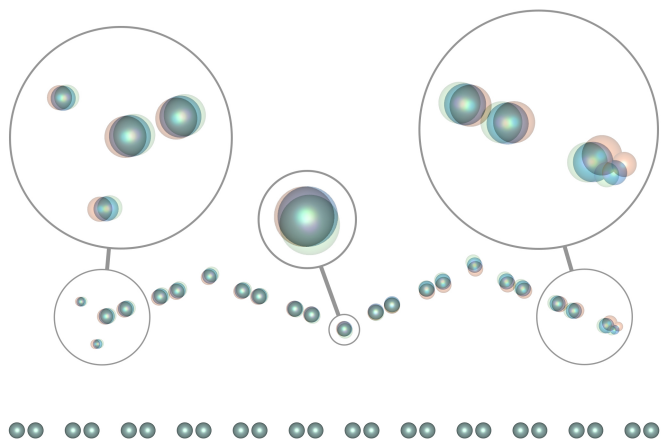

**Figure 20** Overlaid results of the quasi-static motion of an asymmetrical molecule (2H-C16O3-OH) on a graphene substrate. Orange = charge neutral (step 1), green = charging (step 2), and blue = discharging (step 3).

Figure 20 shows the overlay of the GO molecules obtained in the three-step procedure, where insets are the zoom-in views highlighting the atomic positions of the two end chemical groups and the oxygen atom in the center of the molecule. The orange atoms represent equilibrium positions prior to charging. Upon electron injection, the atoms move to new positions depicted by the green spheres. The –H2 end group on the tail and the oxygen atom in the center of the molecule show a motion in the right direction. Meanwhile the –OH end group on the head moves backward slightly (to the left). In this step, the molecule exhibits a contraction as expected. When the injected charges are removed, the molecule elongates. Due to the asymmetric end groups that cause a differential gripping effect with the graphene substrate, the whole molecule exhibits a small displacement to the right, *i.e.*, 0.0021Å. In spite of a very small magnitude, the trend of motion is clear.

Figure 21 summarises the center-of-mass displacements for the three two-unit molecules in one charging/discharging cycle. Electromechanical strains of the free two-unit molecules (*i.e.*, without graphene substrate) are shown for a comparison, revealing a general correlation, whereby a higher electromechanical strain elicits a larger molecule displacement. For a graphene-based actuator in experiments, an AC voltage is supplied in the MHz range [44, 58, 59]. Ideally, an upper limit of the achievable motion could be ~200nm per second for the asymmetric 2H-$C_{16}O_3$-OH molecule. Note that this device is operated based on the quantum mechanical strain from strong covalent bonds. There is low probability of wear, fatigue, and other defects developing during the operation of the robot, promoting a long service life.

We believe that such a proof-of-concept study presented in this section is merely the beginning of graphene oxide's potential in the design of biomimetic robots. It is our hope that these *ab initio* investigation results will stimulate future collaborative studies among theorists and experimentalist in this field.

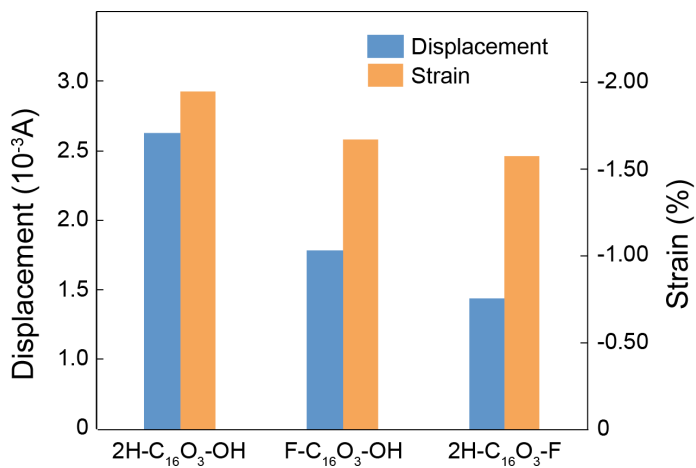

**Figure 21** Linear displacements of three prototype GO biomimetic molecular robots on a graphene substrate in one charge/discharge cycle. The electromechanical strains of the free GO molecules (without substrate) are also shown.

## 5. Conclusions

With a unique combination of two-dimensional molecular structures and superior mechanical and electronic properties, graphene and related materials are ideal candidates for micro- and nano-electromechanical actuators. Since its discovery in 2004, there has been a rapid progress in the research and development of graphene based actuators, with excellent and encouraging results reported. However, compared with other intensive research areas, the study of its actuations is noticeably behind [94]. There is still great potential in utilising this amazing material to address the various demands of actuators in the coming decades. In light of their excellent and distinctive actuation performance characteristics demonstrated through experiments and *ab initio* simulations, graphene oxide could be the next major research focus. Thanks to many available chemical and physical methods [80, 85, 95, 96], the surface molecular structures of graphene can be readily modified to control its physical properties. It is reasonable to anticipate that the vast diversity of graphene oxide molecular structures will bring a rich material database for the design of micro/nano-actuation materials.

Although studies on graphene based electromechanical actuation has attracted much attention,

understanding of the underlying physical mechanisms is still quite poor. *Ab initio* simulation provides a powerful tool to reveal novel actuation physics and helps to stimulate new design ideas. Theoretical models that can predict the actuation performance based on the mechanical, physical, and electrochemical properties of actuation materials are highly desirable as a key to efficiently unlocking novel atomistic based phenomenon. With more in-depth knowledge, we believe that the challenges, such as increasing the strain output, response time, energy conversion efficiency and service life, can be successfully achieved.